\documentclass[11pt]{article}
\usepackage[letterpaper]{geometry}
\newif\ifclean
\cleantrue  
\usepackage{amssymb}
\usepackage{amsfonts}
\usepackage[utf8]{inputenc}
\usepackage[T1]{fontenc}    
\usepackage{graphicx}
\usepackage{subfig}
\graphicspath{{{./}}}
\usepackage{amsmath,bm,amsfonts,amssymb,mathrsfs,mathtools}
\usepackage{color}
\usepackage{hyperref}       
\usepackage{url}            

\usepackage{amsmath,amssymb,amsthm}
\usepackage{bm}
\usepackage{color}
\usepackage[normalem]{ulem}
\newcommand{\COMMENT}[1]{\textcolor{blue}{{[ \sc{#1} ]}}} 

\newcommand{\red}[1]{\textcolor{red}{{#1}}}

\newlength{\figwidth}
\newlength{\figwidthtwo}
\newlength{\figwidththree}
\newcommand{\aref}[1]{App.\,\ref{#1}}
\newcommand{\fref}[1]{Fig.\,\ref{#1}}
\newcommand{\tref}[1]{Table\,\ref{#1}}
\newcommand{\eref}[1]{Eq.\,(\ref{#1})}

\newcommand{\sref}[1]{Sec.\!~\ref{#1}}

\newcommand{\cref}[1]{Ref.\,\cite{#1}}
\newcommand{\crefs}[1]{Refs.\,\cite{#1}}

\newcommand{\ie}{{\it i.e.}\! }
\newcommand{\eg}{{\it e.g.}\! }

\newcommand{\etal}{{\it et al.}\! }

\newcommand{\bs}{\mathsf{b}}

\newcommand{\Ws}{\mathsf{W}}

\newcommand{\Cbb}{\mathbb{C}}

\newcommand{\mub}{{\boldsymbol{\mu}}}

\newcommand{\thetab}{{\boldsymbol{\theta}}}

\newcommand{\epsilonb}{\boldsymbol{\epsilon}}
\newcommand{\sigmab}{\boldsymbol{\sigma}}

\newcommand{\Sigmab}{\boldsymbol{\Sigma}}

\renewcommand{\sb}{\mathbf{s}}

\newcommand{\Vb}{\mathbf{V}}

\newcommand{\Ib}{\mathbf{I}}

\newcommand{\Xb}{\mathbf{X}}

\newcommand{\tr}{\operatorname{tr}}

\newcommand{\dev}{\operatorname{dev}}

\newcommand{\bv}[1]{\mathbf{#1}}

\newcommand{\giv}[1][]{\:#1\vert\:}
\newcommand{\ex}{\mathbb{E}}
\renewcommand{\d}[1]{\ensuremath{\operatorname{d}\!{#1}}}
\DeclarePairedDelimiterX{\infdivx}[2]{(}{)}{%
#1\;\delimsize\|\;#2%
}
\newcommand{\kldiv}{D_{\text{KL}}\infdivx}
\DeclareMathOperator*{\argmin}{arg\,min}
\newcommand{\norm}[1]{\left\lVert#1\right\rVert}
\newcommand{\porosity}{{\varphi}}
\newcommand{\porosityb}{\boldsymbol{\varphi}}
\newcommand{\damage}{\phi}
\newcommand{\damageb}{\boldsymbol{\phi}}
\newcommand{\preddamageb}{\boldsymbol{\phi}_{\rm pred}}
\newcommand{\failure}{{\Phi}}
\newcommand{\strain}{\varepsilon}

\newcommand{\data}{\mathcal{D}}

\newcommand{\NN}{\mathsf{N}\!\mathsf{N}}

\ifclean
\renewcommand{\COMMENT}[1]{{}}
\else
\newcommand{\caution}{\red{\bf Draft} \today \ \red{\bf do not distribute} }
\fi

\title{\bf A heteroencoder architecture for prediction of failure locations in porous metals using variational inference}

\author{
Wyatt Bridgman$^{*,1}$,
Xiaoxuan Zhang\footnote{equal contributions}\,$^{,2}$, \,
Greg Teichert$^2$,\\[0.05in]
Mohammad Khalil$^1$,
Krishna Garikipati$^2$,
Reese Jones$^1$\\[0.1in]
\small
{\it $^1$Sandia National Laboratories, Livermore, CA } \\
\small
{\it $^2$University of Michigan, Ann Arbor, MI }
\normalsize
}

\ifclean
\date{}
\else
\date{\caution}
\fi

\begin{document}

\maketitle

\begin{abstract}
In this work we employ an encoder-decoder convolutional neural network to predict the failure locations of porous metal tension specimens based only on their initial porosities.
The process we model is complex, with a progression from initial void nucleation, to saturation, and ultimately failure.
The objective of predicting failure locations presents an extreme case of class imbalance since most of the material in the specimens do not fail.
In response to this challenge, we develop and demonstrate the effectiveness of data- and loss-based regularization methods.
Since there is considerable sensitivity of the failure location to the particular configuration of voids, we also use variational inference to provide uncertainties for the neural network predictions.
We connect the deterministic and Bayesian convolutional neural networks at a theoretical level to explain how variational inference regularizes the training and predictions.
We demonstrate that the resulting predicted variances are effective in ranking the locations that are most likely to fail in any given specimen.
\end{abstract}


\section{Introduction}

Additive manufacturing (AM) \cite{jared2017additive} and topology optimization \cite{bruns2001topology,le2010stress} have opened a new era for design of complex and optimized components; however, fundamental issues with the porosity of metal AM parts persist.
Experimental assessment tools exist, such as computed tomography (CT) and other non-destructive screening methods, but they need to be coupled to models for quantitative predictions of performance and reliability.
Typically there is large uncertainty in predictions since failure is highly sensitive to small flaws, such as voids, and is, by its nature, an unstable process.
Since failure is usually localized, indications of critical flaws can be somewhat imprecise and yet useful.
Direct simulation of ductile failure is computationally expensive due the complexity of the models and the issues with material stability.
Here we design a convolutional neural network to efficiently predict failure locations given initial porosity at a suitable resolution.
Since the output image is of different character than the input image, as opposed to the widely-employed {\it autoencoder}, the proposed architecture is more accurately described as a {\it heteroencoder} \cite{hachaj2021deep,roweis1999linear}.

Convolutional neural networks (CNNs) \cite{albawi2017understanding} and similar neural networks have been used for a variety of related tasks in physics and engineering.
For instance, Garland \etal \cite{garland2020deep} used a CNN as a screening tool for predicting the mechanical properties of additively manufactured lattices.
Li \etal \cite{li2020reaction} employed an encoder-decoder CNN to predict the state of a reaction-diffusion system.
Nie \etal \cite{nie2018deep} used two different CNN architectures to predict the stress field of cantilever beams.
CNNs have been widely employed in homogenization and related multi-fidelity applications.
Frankel \etal \cite{frankel2019predicting,frankel2020prediction,frankel2022mesh} have employed pixel and graph based convolutional neural networks for homogenization and field predictions.
Teichert and Garikipati \cite{Teichert2018} have used multi-fidelity learning and surrogate optimization as alternates to phase field modeling to predict precipitate morphologies in alloys.
Zhang and Garikipati \cite{Zhang2020Garikipati-CMAME-ML-RVE} also have explored dense neural networks (DNNs) and CNNs with multiresolution learning of the free energy and homogenized the nonlinear elastic response of evolving microstructures.
Teichert \etal \cite{Teichert2019,Teichert2020,teichert2021li} have developed a class of integrable deep neural networks and active learning in the context of scale bridging in materials physics.
Johnson \etal \cite{johnson2022predicting} used a CNN to predict the peak load sustained by simulated tensile specimens with different porosity realizations.
They also demonstrated that the network trained to J2 plasticity response was predictive outside the training conditions.
Zhu and Zabaras \cite{zhu2018bayesian} employed Bayesian CNNs to predict the response of stochastic permeability data.
They show that Stein variational gradient descent can outperform standard variational inference (VI) techniques and Gaussian processes.
CNNs have also been applied in the more general context of solving partial differential equations (PDEs).
Zhang and Garikipati \cite{Zhang2021Garikipati-BNN-weak-solution-PDE-SS} have employed both deterministic and Bayesian neural networks for the direct solution of elliptic PDEs.
Physics informed neural networks (PINNs) \cite{raissi2019physics} were developed as general framework for learning from and predicting field data from PDEs.
Muhammad \etal \cite{muhammad2021machine} used a PINNs-like framework to predict strain fields of experimental tensile specimens using processing and local microstructural information as inputs.
Digital image correlation data was used to provide strains for training.
Particularly related to the present work, Buda \etal \cite{buda2018systematic} discussed and investigated methods to alleviate class imbalance with CNNs.

The contributions of this work to predicting failure with CNNs are manyfold.
We develop a means to ameliorate class imbalance inherent in predicting single/few points of failure in a larger domain through data transformations and loss re-weightings.
We also recast the categorical problem of failure/no-failure at a voxel level to a regression problem and then use precision and recall as scoring metrics that encode the usefulness of predictions.
We relate perturbative sensitivity analysis of the system response to the CNN and the Bayesian CNN (BCNN) predictions.
We connect the formulation of the calibration of deterministic and Bayesian convolutional neural networks at a theoretical level to explain how variational inference regularizes training and prediction.
Furthermore we provide concrete results of the value of using variational inference (VI) with the CNN to provide confidence in the predictions and indications of the epistemic uncertainty of the model.
We then use the predicted mean and variance of the resulting BCNN to effectively rank the likely failure locations.

In \sref{sec:failure} we describe the phenomenology of the failure process of ductile porous metals as represented by a well-calibrated physics-based model.
Then, in \sref{sec:CNN}, we develop the neural network architecture as well as the data transformations and loss manipulations needed for useful predictions.
In \sref{sec:CNN_performance} we demonstrate the performance of the proposed CNN.
Then, in \sref{sec:BCNN}, we augment the CNN architecture with VI and connect the Bayesian augmentation to the deterministic model.
In \sref{sec:BCNN_performance} we illustrate the advantages of the Bayesian network.
We conclude, in \sref{sec:conclusion}, with a summary of our findings and some additional observations in the context of future work.

\section{Ductile failure of porous metals} \label{sec:failure}

To emulate a real failure process we employed a traditional physics-based model calibrated to experimental data consisting of porosity imaging and stress-strain response curves.
The physical model has two levels of porosity: (a) larger pores that are represented explicitly in the mesh, and (b) smaller pores represented implicitly in the viscoplastic constitutive model with damage evolution.
The distinction between the two is the explicit pores were visible with computed tomography (CT) \cite{salzbrenner2017high,boyce2017extreme}, while the implicit porosity was inferred in the calibration process \cite{khalil2019modeling}.
The initial explicit pores $\porosity(\Xb)$, initial background void density $\damage(\Xb,0)$, and loading $\strain(t)$ interact to create an evolving damage field $\damage(\Xb,t)$ and, ultimately, failure locations $\failure(\Xb)$.
As in the high-throughput experiments used to generate the calibration data, we simulate the rectangular gauge section of a tensile specimen loaded to failure.

\subsection{Constitutive Model} \label{sec:material_model}

We employ a previously calibrated model of AM 17-4 PH stainless steel \cite{khalil2019modeling}  to represent the material behavior of the porous tension specimens in this study.
This plasticity model is affected by accumulation of damage that results in local ductile failure and ultimately structural instability with sufficient loading in the 1 mm $\times$ 1 mm $\times$ 4mm  porous tension specimens.

In the constitutive model, stress $\sigmab$ follows a linear elastic rule:
\begin{equation} \label{eq:stress}
\sigmab = (1-\phi) \Cbb ( \epsilonb - \epsilonb_p ) \ ,
\end{equation}
where $\Cbb$ is the isotropic elastic modulus tensor with components which depend on Young's modulus $E$ and Poisson's ratio $v$, $\epsilonb$ is the total strain, and $\epsilonb_p$ is the plastic strain.
The evolution of the void fraction, $\phi$,   effects a reduction of stiffness of the material.

The mechanical behavior evolves according to a coupled set of ordinary differential equations.
The plastic strain ${\epsilonb}_p$ increases according to:
\begin{equation} \label{eq:dotep}
\dot{\epsilonb}_p = \sqrt{\frac{3}{2}} f \sinh^n \left( \frac{\sigma_\text{vm}/(1-\phi) - \kappa}{Y} - 1 \right) \frac{\sb}{\| \sb \|} \ ,
\end{equation}
with ${\epsilonb}_p(t=0) = \mathbf{0}$ and where $Y$ is the yield strength, $\sb = \dev \sigmab$ is the deviatoric stress, and $\sigma_\text{vm} \equiv \sqrt{\frac{3}{2} \sb\cdot\sb}$ is the von Mises stress.
The evolution of the isotropic hardening variable $\kappa$ is governed by:
\begin{equation} \label{eq:hardening}
\dot{\kappa} = (H-R \kappa) \, \dot{\epsilon}_p \ ,
\end{equation}
with $\kappa(t=0) = \kappa_0$ and where $H$ is the hardening modulus, $R$ is the recovery coefficient, and $\dot{\epsilon}_p = \sqrt{\frac{2}{3} \dot{\epsilonb}_p \cdot \dot{\epsilonb}_p}$ is the equivalent plastic strain rate.
The damage due to implicit void volume fraction $\phi$ evolves according to:
\begin{equation} \label{eq:phidot}
\dot{\phi} = \sqrt{\frac{2}{3}}\dot{\epsilon}_\text{p}\frac{1 - (1-\phi)^{m+1}}{(1-\phi)^{m}}\sinh\left(\frac{2(2m-1)}{2m+1}\frac{p}{\sigma_\text{vm}} \right) + (1-\phi)^2 \dot{\eta}\nu_0 \ ,
\end{equation}
which includes nucleation and growth of voids, where the damage exponent $m$, and the volume of newly nucleated voids $\nu_0$, are additional model parameters.
To model nucleation, the void concentration $\eta$, evolves according to:
\begin{equation} \label{eq:etadot}
\dot{\eta} = \left( N_1 \left(\frac{2^2}{3^3} - \frac{J_3^2}{J_2^3}\right) + N_3 \frac{p}{\sigma_\text{vm}} \right) \eta \, \dot{\epsilon}_p
\end{equation}
where $p = 1/3 \, \sigmab\cdot\Ib$ is the pressure, $J_2 = 1/2 \tr \sb^2 = 1/3 \, \sigma_\text{vm}^2$ and $J_3 = 1/3 \tr \sb^3$.
Lastly, once the void fraction $\phi$ exceeds a threshold $\phi_\text{max}$ the material is considered completely failed and acts as an explicitly represented pore.

The calibrated parameters are given in \tref{tab:parameters} and additional details can be found in \cref{khalil2019modeling}.

\begin{table}[t]
\centering
\begin{tabular}{|lc|c|}
\hline
Parameter & & Value \\
\hline
Young's modulus (GPa)         & $E$       & 240 \\
Poisson's ratio               & $v$       & 0.27 \\
Yield strength (MPa)          & $Y$       & 600   \\
Initial void size ($\mu$m$^3$)    & $\nu_0$   & 0.1 \\
Initial void density ($\mu$m$^{-3}$) & $\eta_0$  & 0.001  \\
Flow exponent                 & $n$        & 10 \\
Damage exponent               & $m$        & 2 \\
Flow coefficient              & $f$        & 10        \\
Isotropic dynamic recovery    & $R$        & 4\\
Isotropic hardening (GPa)     & $H$        & 5\\
Initial hardening (MPa)       & $\kappa_0$ & 460\\
Initial damage                & $\phi_0$   & 0.08\\
Shear nucleation              & $N_1$      & 10      \\
Triaxiality nucleation        & $N_3$      & 13\\
Maximum damage                & $\phi_\text{max}$  & 0.5  \\
\hline
\end{tabular}
\caption{Constitutive model parameters from \cref{khalil2019modeling}.
}
\label{tab:parameters}
\end{table}

\subsection{Generation of porosity realizations} \label{sec:KLE}
As in our previous work \cite{khalil2019modeling} we utilized a 20$\times$20$\times$80 structured mesh of 8 node hexahedral elements.
On this grid we generated the initial explicit porosity field via a Karhunen–Lo\`{e}ve (KL) process with $\approx$ 12,000 modes calibrated to the average porosity $\bar{\porosity} = 0.0008$ and spatial power-exponential correlation function fitted to the empirical correlation information as extracted from the available CT scans.
Since the observed porosity is confined to clusters that consist of only a few voxels, the correlation length for the porosity process is short.
Hence many modes are needed to resolve the high frequencies and capture 99.9\% of the energy.
Full details are given in \cref{khalil2019modeling}.

\subsection{Boundary value problem}

Tensile loading was imposed with minimal displacement boundary conditions on the top (high $z$) and bottom (low $z$) surfaces of the specimen such that these surfaces remained planar but could change dimension laterally.
A nominally quasi-static strain rate, $\dot{\strain}$ = 0.002/s, was applied through these boundary conditions until structural failure.

\subsection{Phenomenology of the failure process}
\label{sec:phenomenology_failure_process}
\begin{figure}[h!]
\centering
\includegraphics[width=0.90\linewidth]{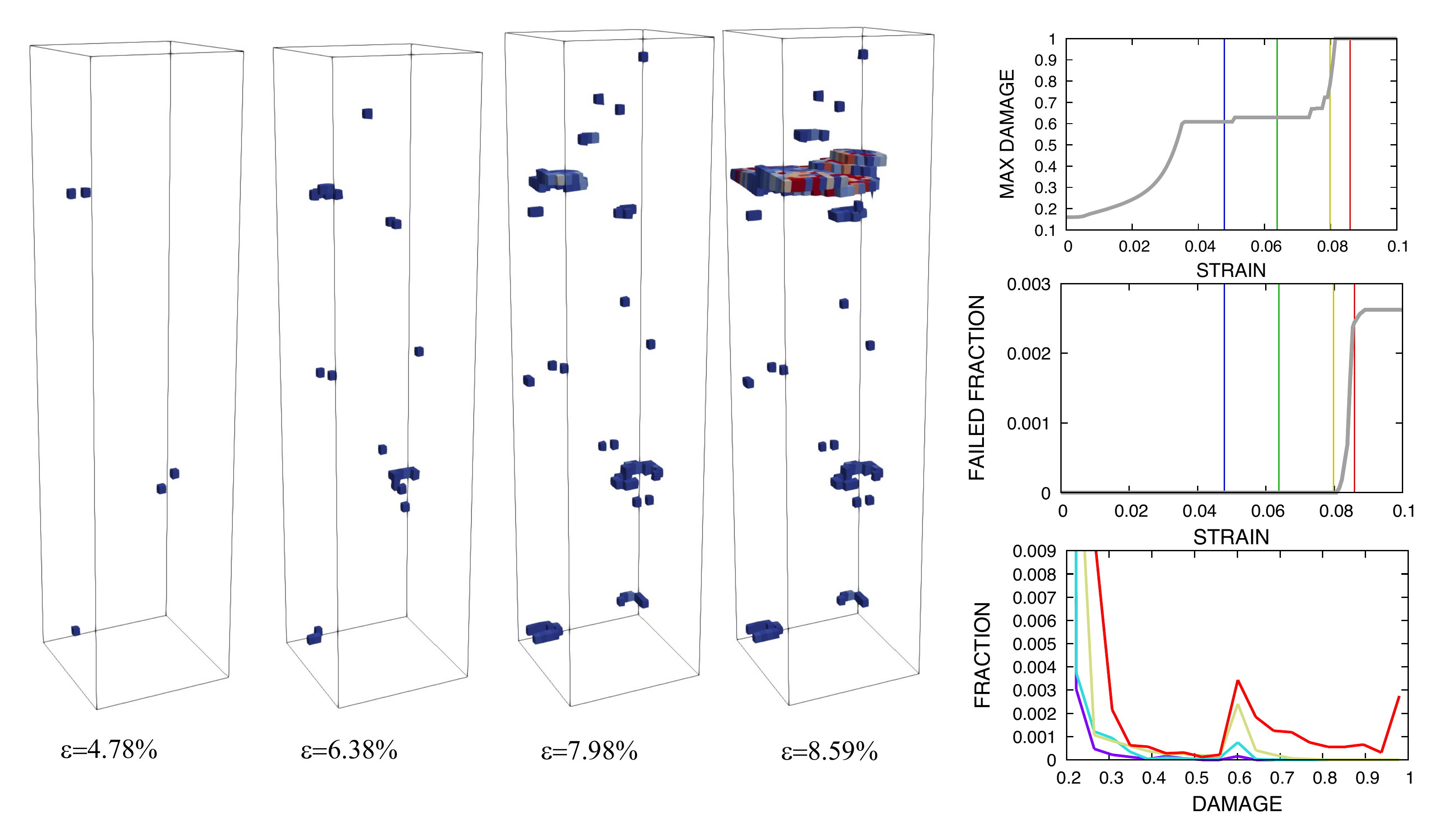}
\caption{Representative damage evolution example.
For the damage field sequence (left) blue:$\damage/\damage_\text{max} = 0.6$, red: $\damage/\damage_\text{max} = 1.0$. Colored vertical lines in the history plots (right) correspond to strain levels of the damage histograms (right, bottom) and the damage field sequence (left).}
\label{fig:damage_example}
\end{figure}

\begin{figure}[h!]
\centering
\includegraphics[width=0.90\linewidth]{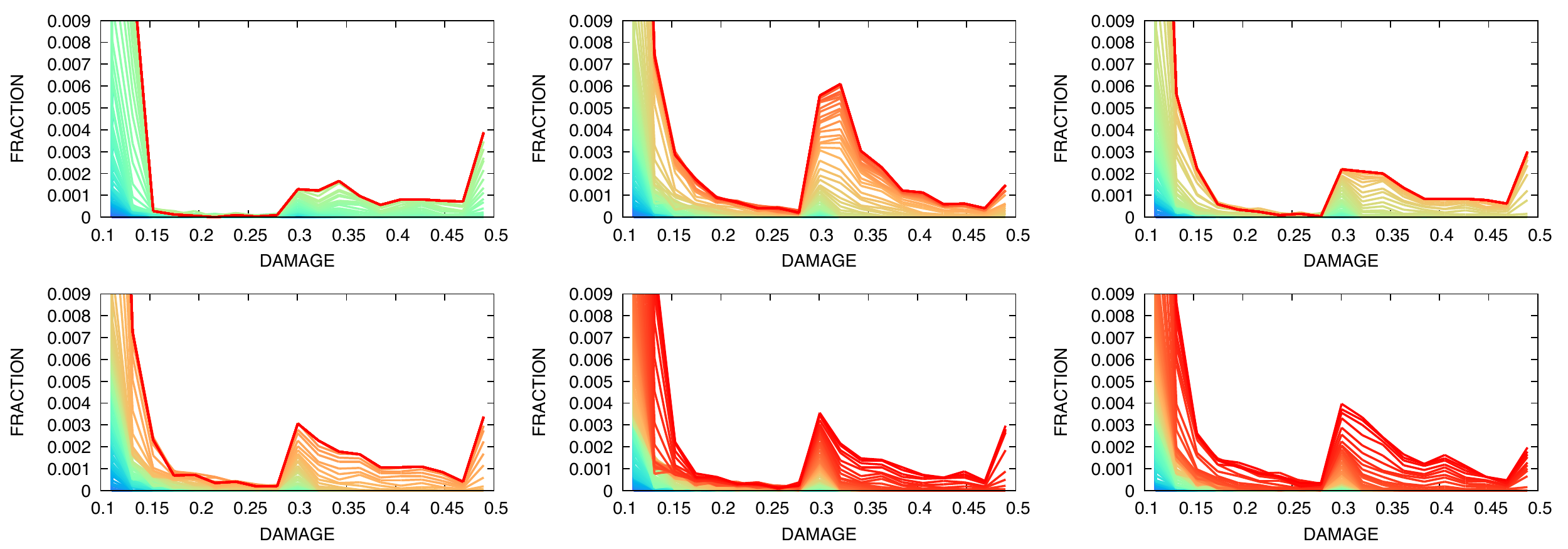}
\caption{Representative damage evolution histograms of six selected realizations.
Blue: $\strain = 0$, red: $\strain \approx$ 9 \%.
Damage is reported as the fraction: $\phi/\phi_\text{max}$.}
\label{fig:damage_histograms}
\end{figure}

Damage evolution leading to failure progresses through three stages in these specimens.
\fref{fig:damage_example} shows the response of a representative sample.
In stage I ($\strain \approx$ [0:2]\%) there is a smooth rise of maximum damage in the sample from the background  $\porosity_0$=0.008 with no material failure.
In stage II ($\strain \approx$ [2:4]\%) there is a plateau of maximum damage at $\damage/\damage_\text{max} \approx 0.6$ as sites nucleate damage, harden and then alternate sites start to damage.
This stage ends when all the likely/vulnerable sites saturate.
In stage III ($\strain \approx$ [4:5]\%) a single most vulnerable site (or in some rare cases a few clustered sites) reaches failure and the specimen quickly fails with a horizontal ductile fracture surface.
This progression has some qualitative similarity with the well-known phenomenon of the evolution of L\"{u}der bands \cite{mesarovic1995dynamic}.
The location along the tensile axis, \ie, the $z$-location, of the first element(s) to reach failure is highly correlated with the ultimate $z$-location of failure as measured at the final load step.
In fact, the difference in $z$-location of the first and final failure is on average only about 0.04 mm, or 1\% of the specimen length or about one element, which indicates the first elements to fail are the most likely cause of ultimate failure.
The histograms of damage for a sequence of strains are shown in \fref{fig:damage_histograms}.
The panels of \fref{fig:damage_histograms} for different realizations illustrate that the three stage progression is representative of the ensemble.

\begin{figure}[h!]
\centering
\includegraphics[width=0.45\linewidth]{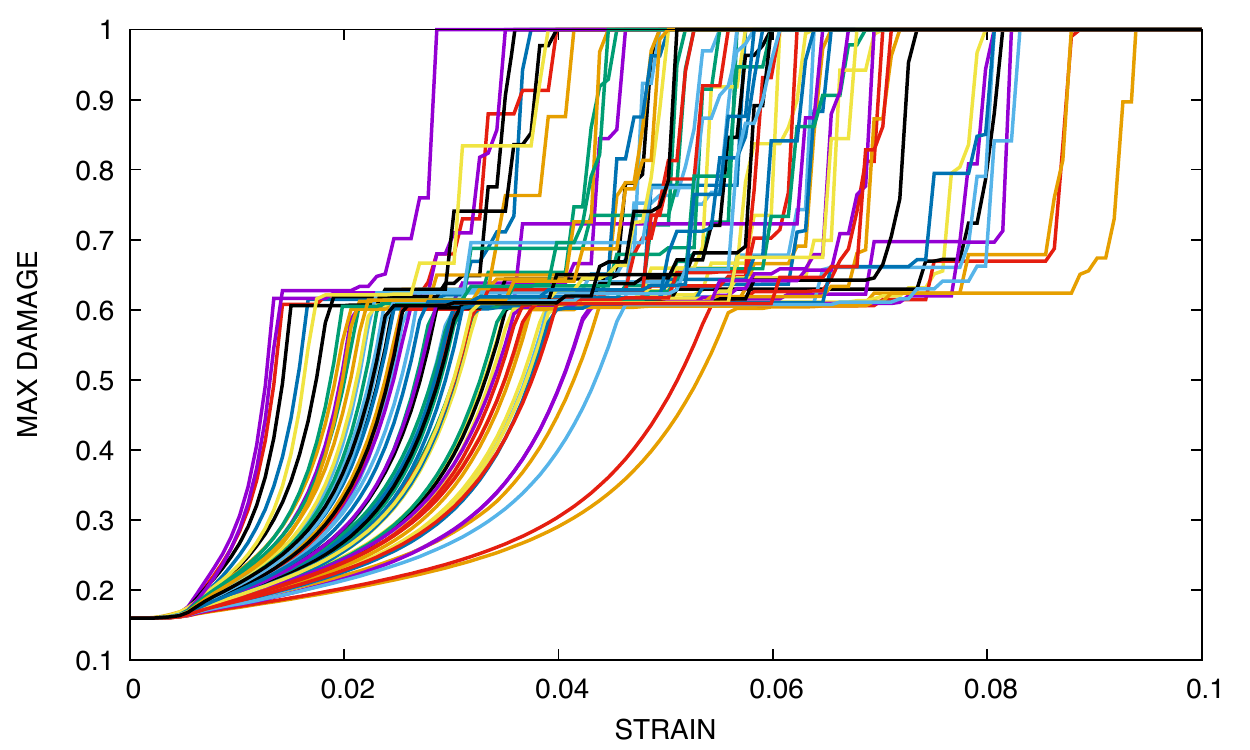}
\includegraphics[width=0.45\linewidth]{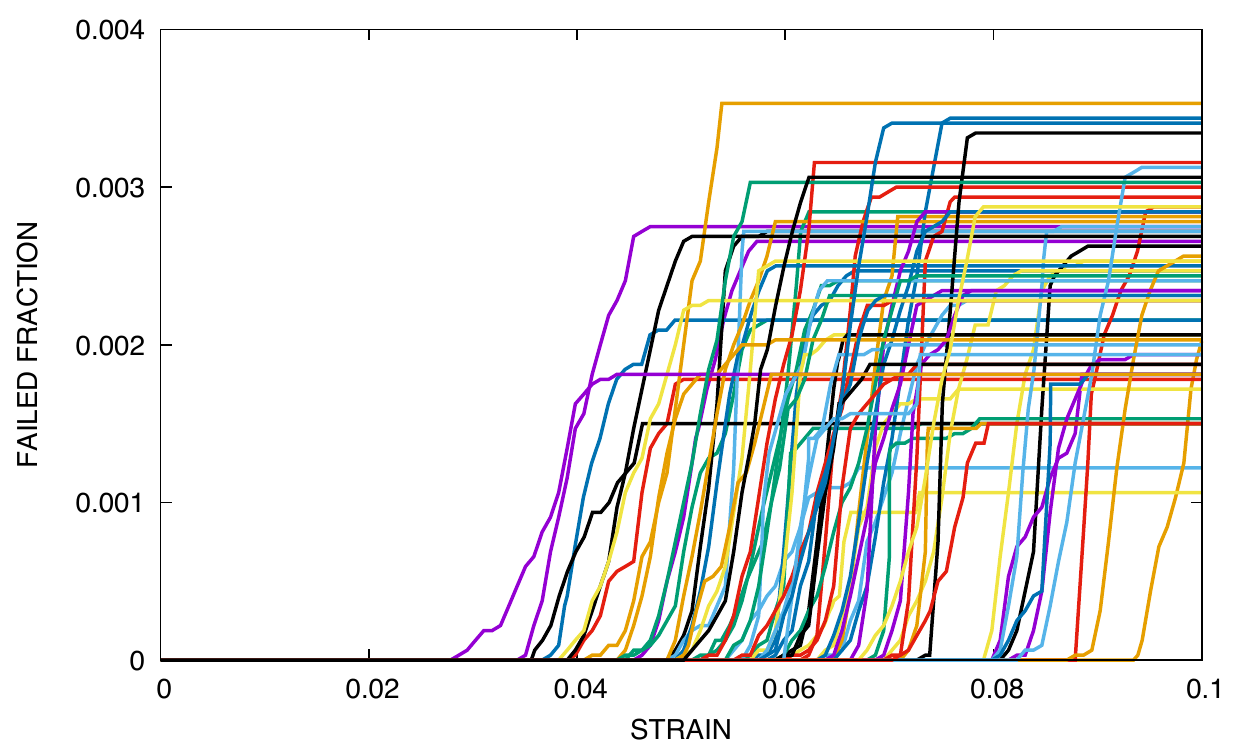}
\caption{Ensemble response with strain: (a) maximum damage, and (b) failed volume fraction.
Different realizations are distinguished by different color curves.
Damage is reported as $\damage/\damage_\text{max}$.}
\label{fig:ensemble_response}
\end{figure}

Lastly, the bundle of response curves shown in  \fref{fig:ensemble_response} demonstrates that there is considerable variability within the ensemble.
The progression of maximum damage in the samples over time is similar, but the rise, plateaus, and final failures vary considerably with the different pore configurations.
Also once an element fails (at $\phi/\phi_\text{max}=1$ locally) the rise to structural failure is rapid for all samples but the strain at which it occurs, and the void volume fraction, $\phi$, at which the structure ultimately fails vary substantially.

\subsection{Sensitivity of failure to pore locations}

\begin{figure}[h!]
\centering
\includegraphics[width=1.0\linewidth]{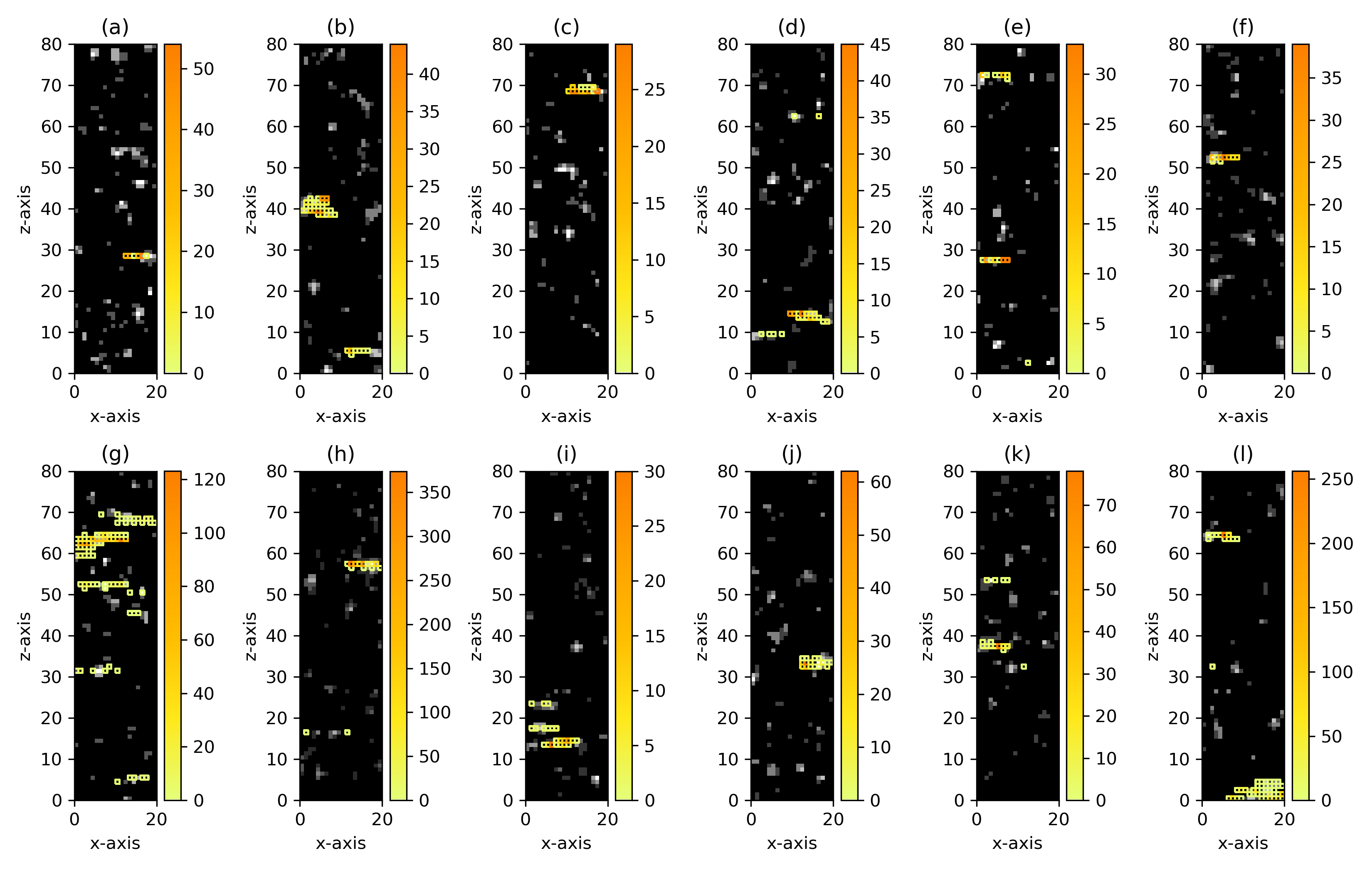}
\caption{Sensitivity analysis results for twelve realizations, where the upper row corresponds to the realizations in \fref{fig:damage_histograms}.
The grayscale background represents the number of void voxels summed along the y-axis in the original structure, where black represents zero pores.
The colored squares mark where failure has occurred  in the sensitivity analysis; the color indicates the number of occurrences.
}
\label{fig:sa-results}
\end{figure}

Although the damage criterion used to define the failure location typically provides a single region of failure, there are multiple regions of high damage that compete in early stages of the loading (stage II).
This suggests that small variations in the explicit porosity could lead to failure occurring in alternate regions of the specimen.
To explore this sensitivity, we carry out one voxel perturbations of the porosity field.
We randomly select a void voxel on the edge of a pore within a given distance from the original failure location and swap it with a randomly selected solid voxel anywhere in the specimen.
This process slightly decreases the size of the pore in the neighborhood of original failure while maintaining the same overall porosity, but potentially alters the porous surface area.
Loading the modified porosity structure results in a failure location that may or may not be close to the failure location of original structure which reflects the sensitivity of the failure location to the porosity field.
This process was repeated 100 times for nine realizations and 500 times for three realizations (realizations (g), (h), and (l) in  \fref{fig:sa-results}).

The locus of failure locations are shown along with the explicit porosity of the unperturbed realization in  \fref{fig:sa-results}.
The porosity is projected onto the $x$-$z$-plane by summing the number of voided voxels along the $y$-axis and plotting the result as a grayscale value, where black indicates no pores.
The failure locations are shown using colored squares, where multiple failures with the same $x$ and $z$ values are indicated using the color scale next to each plot.
Some realizations, such as (a), (c), and (f) exhibit very little sensitivity to the random perturbation of a single void voxel.
Others, \eg (b), (i), (k), demonstrate two or three regions of failure, while realization (g) displays the most sensitivity, with failure occurring at several different locations within the specimen.

\section{A convolutional neural network model of localized failure} \label{sec:CNN}

Our goal in using neural networks is to predict likely failure locations, which is a binary categorization $\failure(\Xb)$ (failed/not failed), given the binary initial porosity field $\porosity(\Xb)$.
Initial failure is defined as the first time at which the damage $\damage(\Xb)$ in any element/voxel at $\Xb$ reaches $\damage_\text{max}$, at which point the element will start to act as a void and initiate structural instability.
The 20$\times$20$\times$80 grid that discretizes the domain makes this a high, yet finite, dimensional problem.
The emphasis on predicting the failure locations leads to stark class imbalance, where the failed to not failed ratio is on the order of 1:10$^4$, since only a few of the elements fail in any particular mesh.
The fact that a prediction of no failure anywhere scores nearly as well in mean error as a perfect prediction if there are only a few failures illustrates the class imbalance problem.
So even relative accuracies of 0.1--0.01 \% would be insufficient to reliably predict one element failing in a mesh of 32,000 elements.

As a first step to alleviating this issue, we recast this categorical prediction problem into a regression problem of predicting the underlying damage field $\damage(\Xb,t_\text{fail})$ at time of failure $t_\text{fail}$.
Although the damage fields are still non-smooth, the spatial correlations in the damage field make the regression task easier; in effect this recasting is a physically motivated regularization.
Recall, the physics-based direct numerical simulations generate training data consisting of an input binary porosity field $\porosity(\Xb)$  mapped to an output damage field $\damage(\Xb,t_\text{fail})$ at the time of first failure $t_\text{fail}$.
Since the transition of $\damage \to \damage_\text{max}$ in a particular location rapidly leads to complete structural failure, we have a natural, direct failure definition and a binary failure location field $\failure(\Xb)$ (output).
So a high dimensional, two category, $\failure \in \{0,1\}$, problem with data $\data=\{ \porosity, \failure \}$ has a direct connection to the regression problem with $\data=\{ \porosity, \damage \}$.
To obtain a dataset $\data$ with 110,712 full-field instances, we generated 18,452 realizations and augmented this ensemble with 6-fold symmetry transforms ($\pi/2$ rotations about the tensile axis and reflections in each of the cardinal directions).
In addition to the regularization of recasting the categorical problem, predicting a continuous, albeit irregular, field in the regression problem allows for data manipulation (\sref{sec:data_transformation}) and loss re-weighting (\sref{sec:weighted-loss}) to further regularize the problem and promote accurate prediction of the failure locations.
In this mode, failure locations are recovered post-prediction by applying a threshold to the predicted damage field.

\subsection{Data transformations} \label{sec:data_transformation}

As can be seen in Figs. \ref{fig:damage_example} and \ref{fig:sa-results} the damage fields can have isolated high damage voxels as well as more significant clusters that correlate with the ultimate failure locations.
Based on this observation, we applied smoothing, contrast, and normalization transformations to condition the data and improve the CNN training without sacrificing the ability to predict failure locations.

We first applied a Gaussian filter to the damage field $\damage$ to obtain:
\begin{equation} \label{eq:filter}
\damage' = \text{GF}(\damage, k_0) = g \ast \damage \quad \text{with} \quad g = \frac{1}{(2\pi k_0)^{3/2}} \exp\left( -\frac{\| \Xb \|^2}{2 k_0^2} \right),
\end{equation}
with $k_0$ being the width of Gaussian kernel $g$, $\| \Xb \|$ being the distance between the current filter location and a neighboring location, and $\ast$ being convolution.
The use of a spatial smoothing filter was motivated by the fact that damage fields in metals are typically more smooth and regularly distributed than the direct numerical simulation results, since the viscoplastic model in \sref{sec:material_model} lacks an intrinsic length-scale.
Furthermore, convolution such as \eref{eq:filter} has the effect of removing high frequencies, which is compatible with the encoder-decoder network (\sref{sec:heteroencoder}) we employed that distills the full resolution images to a reduced dimensionality latent space.
It is also consistent with the fact that the KL process (\sref{sec:KLE}) that generated the porosity realizations is dominated by the low frequency components.
For this dataset $k_0<1.0$ mm had a negligible smoothing effect, whereas $k_0>2.0$ mm removed important features (clusters of high damage); hence,  kernel width $k_0=1.5$ mm provided a good balance between the two effects.

Next, we applied the $\operatorname{softmax}$ function voxel-wise to each voxel $I$ in the damage field $\damage'$ of a realization:
\begin{equation}
\tilde{\damage}_I = \operatorname{softmax}( \damage', c_0)
= \frac{\exp{(c_0 \damage'_I)} }{\sum_J \exp(c_0 \damage'_J)}
\quad \forall\ I,
\end{equation}
with $c_0$ being the contrast parameter that increases the difference between low and high values.
The $\operatorname{softmax}$ transformation is commonly used in categorization and image processing.
In our application, a larger $c_0$ results in higher contrast between high and low damage values, \ie bringing the damage field closer to a binary field, and emphases the clusters of high damage.
We used $c_0=5.0$ based on preliminary studies.
Note that that the contrast and smoothing parameters could be treated as hyperparameters in the CNN training but the fact that extremes in their values lead to trivial predictions presents challenges which are left for future work.

Finally, we normalize the resulting damage field $\tilde{\damage}$ per realization to obtain:
\begin{equation}
\bar{\damage}_I = \frac{\tilde{\damage}_I - \min_j \tilde{\damage}_J}{\max_J \tilde{\damage}_J - \min_J \tilde{\damage}_J } \quad \forall\ I.
\end{equation}

\begin{figure}[h!]
\centering
\includegraphics[width=0.9\linewidth]{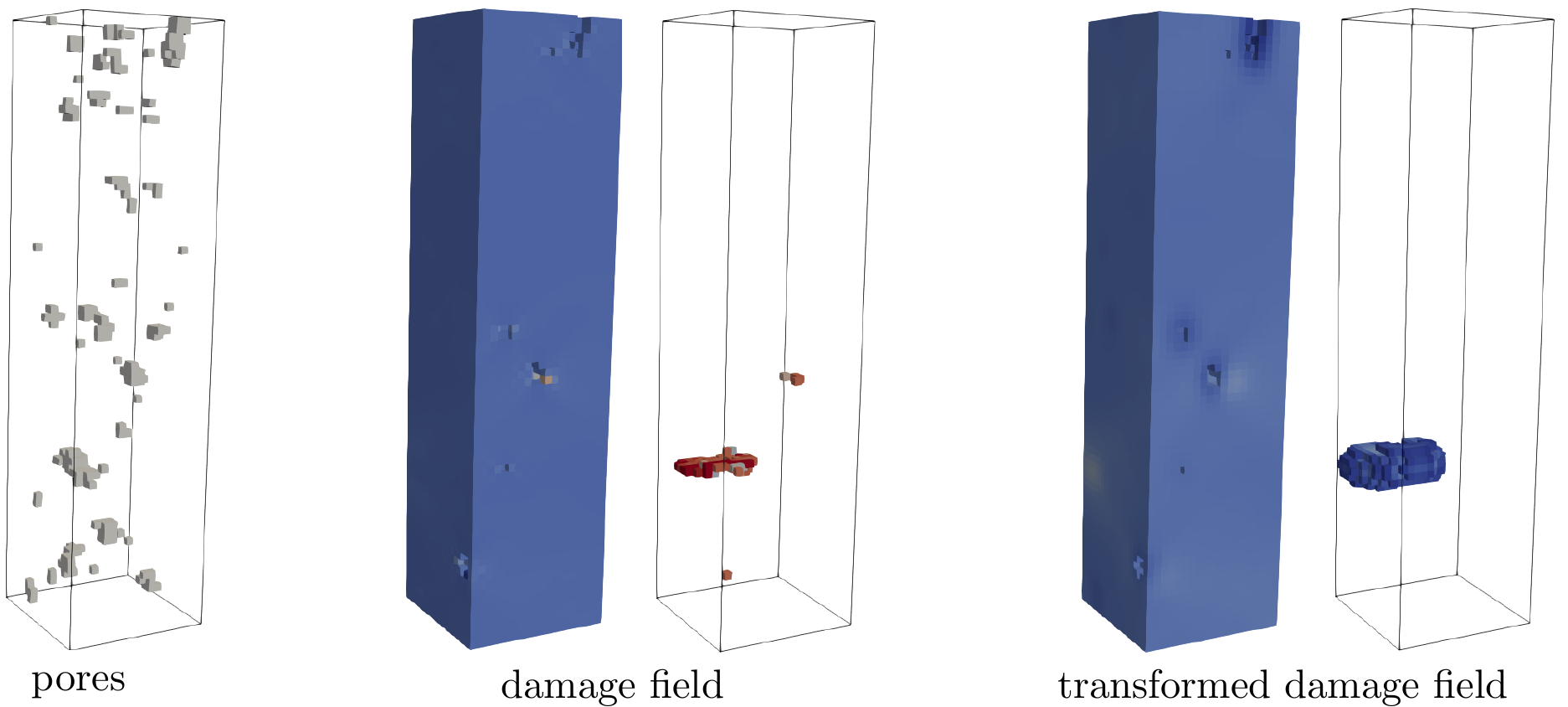}
\caption{Illustration of the transformation of a representative porosity field.
Left: the initial pore distribution.
Middle: the damage field from the direct numerical simulation and the locations of damage where $\damage/\damage_\text{max} \ge 0.6$.
Right: the transformed damage field after Gaussian filtering with a kernel width $k_0=1.5$ mm and the $\operatorname{softmax}$ transformation with a contrast parameter $c_0=5.0$, and the locations of the smoothed damage field where $\damage/\bar{\damage}_\text{max} \ge 0.6$.
Note the cluster shown on the right encloses voxels with higher values of damage.}
\label{fig:data-transformation}
\end{figure}

As discussed in \sref{sec:phenomenology_failure_process} and shown in \fref{fig:damage_example}, initially damage can rise at multiple sites due to nucleation and hardening, but ultimately only a most vulnerable site in a given porosity field will reach failure.
Failure typically occurs at one or a few voxels with the maximum damage value $\damage_\text{max}$ which are surrounded by a cluster of high damage.
Since Gaussian smoothing results in a weighted damage value that accounts for the values in the surrounding neighborhood, ideally smoothing will: (a) remove isolated damage voxels that developed in stage II but did not lead to ultimate failure in stage III and (b) cluster the voxels surrounding the voxel that ultimately fails.
The contrast transform operates locally but also suppresses low damage values and enhances high values.
As shown in \fref{fig:data-transformation},
the smoothing treatment emphasizes the region where failure ultimately occurs in stage III.
The effectiveness of the different data transformations in promoting training and accurate predictions is discussed  \sref{sec:cross-validation}.

\subsection{Weighted loss functions} \label{sec:weighted-loss}

As stated, predicting failure voxel locations with a convolutional neural network from porosity realizations can be naturally recast as a damage field regression problem.
Hence, we use the mean squared error (MSE) of the predicted damage field as the optimization objective function for the regression problem:
\begin{equation} \label{eq:MSE_loss}
\frac{1}{N_v N_s} \sum_{s=1}^{N_s} \|\damageb_{s} - \hat{\damageb}_{s}\|^2
\end{equation}
where $\damageb = \{ \damage_I \}$ is a vector of damage values at voxels for a specific sample, $N_s$ is the number of samples in the training set, $N_v$ the number of voxels in a damage field realization, and $\damageb_s$, $\hat{\damageb}_s$ are the true and predicted fields on the grid for sample $s$, respectively.
Since the damage field at the pore locations is not defined, we omit voxel locations corresponding to the initial voids in \eref{eq:MSE_loss}.

\begin{figure}[h!]
\centering
\includegraphics[width=0.9\linewidth]{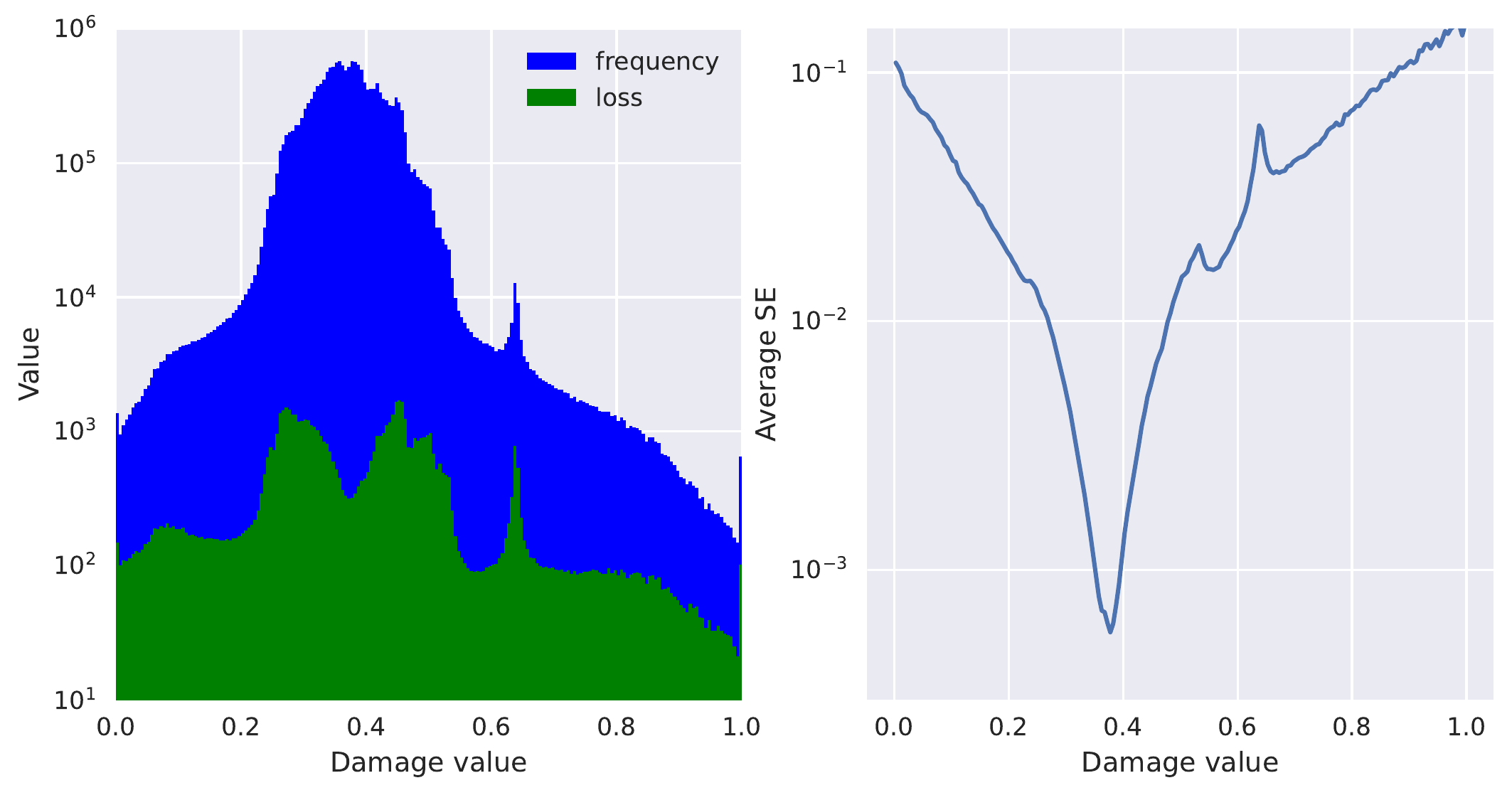}
\caption{Histograms of damage values and MSE loss components partitioned according to the true damage value (left).
Frequency-normalized MSE loss distribution reflecting the average squared error (SE) at each damage value (right).}
\label{fig:damage-value-distribution}
\end{figure}

A primary issue with this standard loss function in the present context is that the damage fields are unbalanced in terms of the number of voxels with small damage values versus the number of voxels where there is significant damage.
This discrepancy causes a CNN trained to this loss to learn the mapping for the low damage regions accurately at the expense of the high damage regions which are of greater importance.
The distribution of voxels and regression errors  depicted as a function of damage value in \fref{fig:damage-value-distribution}a demonstrate this.

The frequency normalized error of the trained CNN predictions in \fref{fig:damage-value-distribution}b illustrate the effect damage value frequency has on the solution found by the optimizer.
Gradient descent finds a local minimum where the reconstruction error is small around the highest frequency damage value and increases quickly away from this value.
This can be understood by partitioning the terms of the MSE loss into two bins according to the true damage voxel value:
\begin{equation}
\frac{1}{N_v N_s} \left [ \sum_{\damageb_s,\hat{\damageb_s} \in \mathcal{D_{\text{low}}}}\|\damageb_s - \hat{\damageb}_s\|^2 +                                  \sum_{\damageb_l,\hat{\damageb_l} \in \mathcal{D_{\text{high}}}}\|\damageb_l - \hat{\damageb}_l\|^2 \right] \ ,
\end{equation}
where $|\data_{\text{low}}| \gg |\data_{\text{high}}|$ and $\data_{\text{low}}$ contains all the relatively undamaged voxels (those below a threshold $\approx \damage_\text{max}$) and $\data_{\text{high}}$ the few voxels that approach failure at $\damage_\text{max}$.
Hence, given relatively uniform errors across each sample, the contribution to the gradient of the predominant $\data_{\text{low}}$ data is likely to be significantly larger than the contribution of $\data_{\text{high}}$.
This leads the optimizer to ``overlearn'' the high frequency low damage values.

\begin{figure}[h!]
\centering
\includegraphics[width=1.0\linewidth]{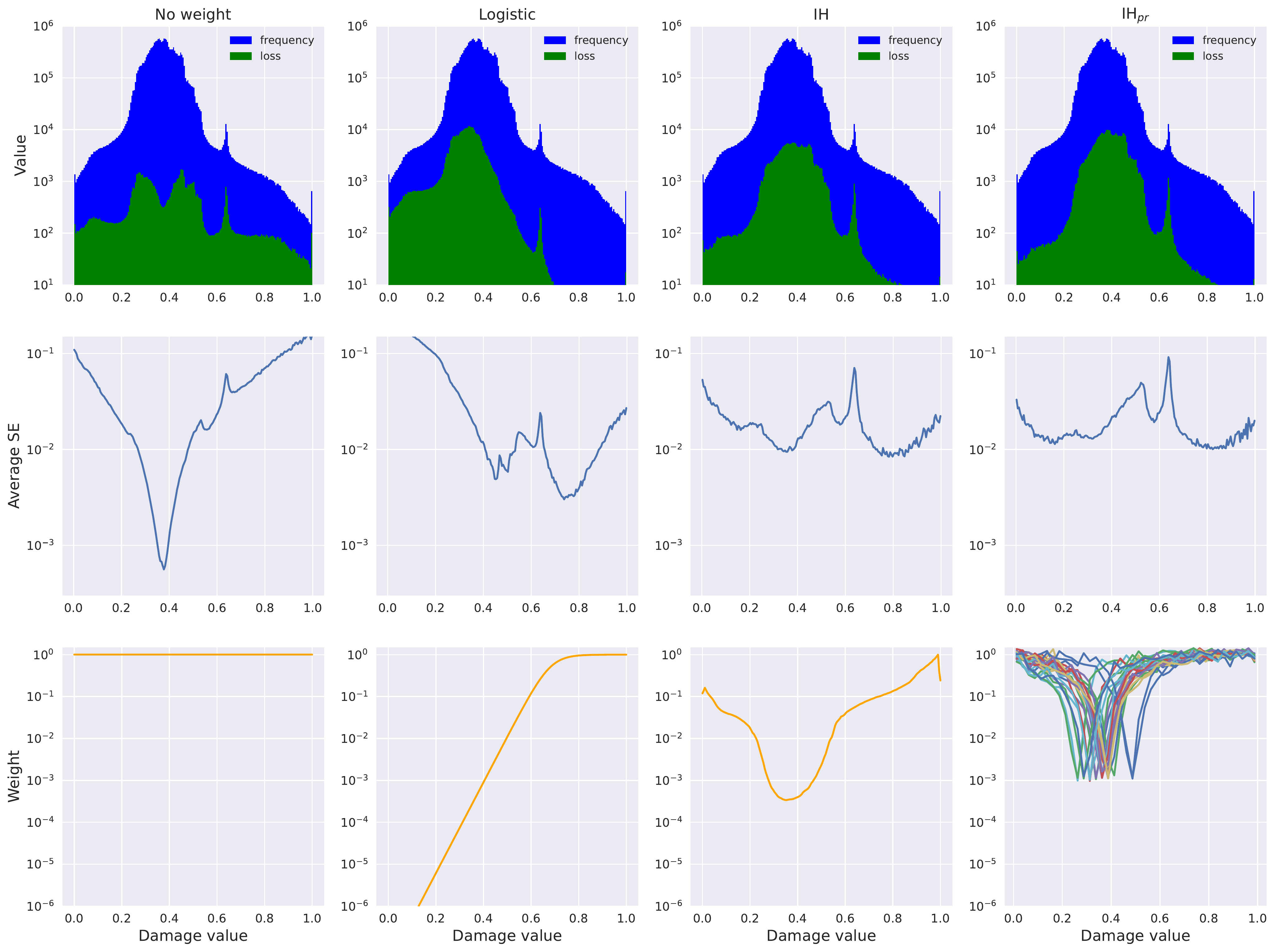}
\caption{Response of optimization to different weight functions: uniform weighting (no weight), logistic, inverse histogram (IH), and inverse histogram per realization ($\text{IH}_{pr}$), left to right.}
\label{fig:opt-response-to-cost-function-weights}
\end{figure}

This issue in our regression problem is related to class-imbalance in classification problems \cite{japkowicz2002class,guo2008class,abd2013review,buda2018systematic,johnson2019survey} where training data is unevenly distributed among the classes leading to solutions which are inaccurate for classes with fewer training examples.
Common techniques to address this issue are loss-based techniques, where the loss function is modified with a weighting scheme to balance contributions from different classes \cite{buda2018systematic}, and data-based techniques, where sampling strategies are used to balance the training data \cite{buda2018systematic}.
In this work, we use the loss weighting approach as the imbalance exists in each field realization so that common sampling methods would not be effective.
Weightings of the MSE loss take the form:
\begin{equation}
\frac{1}{N_v N_s} \sum_{s=1}^{N_s} \| \damageb_{s} - \hat{\damageb}_{s}\|^2 \, w(\damageb_{s}) \ ,
\end{equation}
where $w(\damageb)$ is the per-voxel weighting function which depends only on the true damage values $\damageb$.
Several possible weighting functions are depicted in \fref{fig:opt-response-to-cost-function-weights}: uniform weighting, logistic weighting, inverse histogram (IH) weighting based on the ensemble, and inverse histogram per realization ($\text{IH}_{pr}$) weighting.

The optimization procedure responds to the weighting by finding a solution with a different distribution of the accuracy per damage value shown in \fref{fig:opt-response-to-cost-function-weights}.
By weighting the high-frequency true values with smaller weights, the regression accuracy can be made more uniform across the range of damage values.
While using a weighting inversely proportional to the damage value frequency, either per realization or for the whole ensemble, seems to create the most even distribution of prediction accuracy, the logistic weighting provides a similar effect on the loss.
A comparable effective treatment to inverse histogram weighting could likely be achieved by optimal choice of the logistic functions parameters.
To avoid this parameter tuning, we focus on the inverse histogram weightings which are generated directly from the training data.
Note that a drawback of reweighting is that by down-weighting the contribution of smaller damage values, the loss function penalizes poor small damage value predictions less which leads to more false positives.
Although predicting the low background damage is not the primary goal, this behavior is a consideration in choosing the optimal weight function $w$.
Related discussion will be given in \sref{sec:false_positives}.

\subsection{A heteroencoder convolutional neural network} \label{sec:heteroencoder}

In \sref{sec:KLE} we described a reduced-dimensional representation of the explicit porosity process based on a Karhunen-Lo\`{e}ve expansion (KLE).
It is optimal in the sense that given a fixed number of modes, no other linear decomposition can capture as much energy of the random porosity process as the KL expansion.
In this setting, we decided to capture 99.9\% of the energy which requires $\approx$12,000 KLE modes.
In other terms, the random porosity process can be optimally captured using $\approx$12,000 parameters, namely coefficients of orthogonal modes of a linear decomposition.
In the context of our modeling challenge, KLE, being a linear approximation, can be used to project the input signal onto $\approx$12,000 vectors.
This is a form of encoding of the high-dimensional porosity realizations which serve as inputs to regression problem.

Our problem could be modeled with a linear encoder-decoder
\begin{equation}
\damageb = \Ws_2 \Ws_1 \porosityb + \bs
\end{equation}
where $\Ws_1$ is an encoding linear transform/matrix, $\Ws_2$ is a decoding one and $\bs$ is a bias correction.
The KL process already provides the optimal version of $\Ws_1$, being the collection of all ``linear'' KLE modes.
However, this is not an auto-encoder since the output (damage field, $\damage$) is not an approximation of the input (porosity field, $\porosity$).
This linear model, which we call a heteroencoder \cite{roweis1999linear}, attempts to reduce the input porosity to a smaller latent space then expands back to the high-dimensional output, namely the damage field.
The reduction is formally achievable given the dimensionality of the KLE approximation of the porosity process used to generate the input $\porosity$.

Starting with the KLE as a baseline for dimensionality reduction from the na\"ive dimensionality of the porosity on the grid, the dimensionality of the latent space can be further reduced using a non-linear latent variable model.
The idea is similar to the use of non-linear principal component analysis (PCA) methods, and uses neural network training procedures to generate nonlinear features to achieve further dimensionality reduction of the latent variable space \cite{kramer1991nonlinear}.
This can be enacted using networks with nonlinear activation functions.

In general, an output field often pertains to a small number of functionals of the input process (or latent variables thereof) \cite{tipireddy2014basis}, which justifies further reduction of the number of requisite latent variables.
In our context, the output field to be modeled relates to regions of high damage near failed elements which should depend on a subset of, rather than all, the latent variables.
The convolutional architecture and the spatial smoothing of the input porosity process (achieved in a data pre-processing step) helps further reduce the requisite number of latent variables by diminishing high-frequency content in the input.
Lastly, we note that the use of softmax to define the voxel-wise vector of damage field values, $\tilde{\phi}_I$, confers an attention network form to the overarching framework \cite{graves2016hybrid}.

\begin{figure}[h!]
\centering
\includegraphics[width=1.0\linewidth]{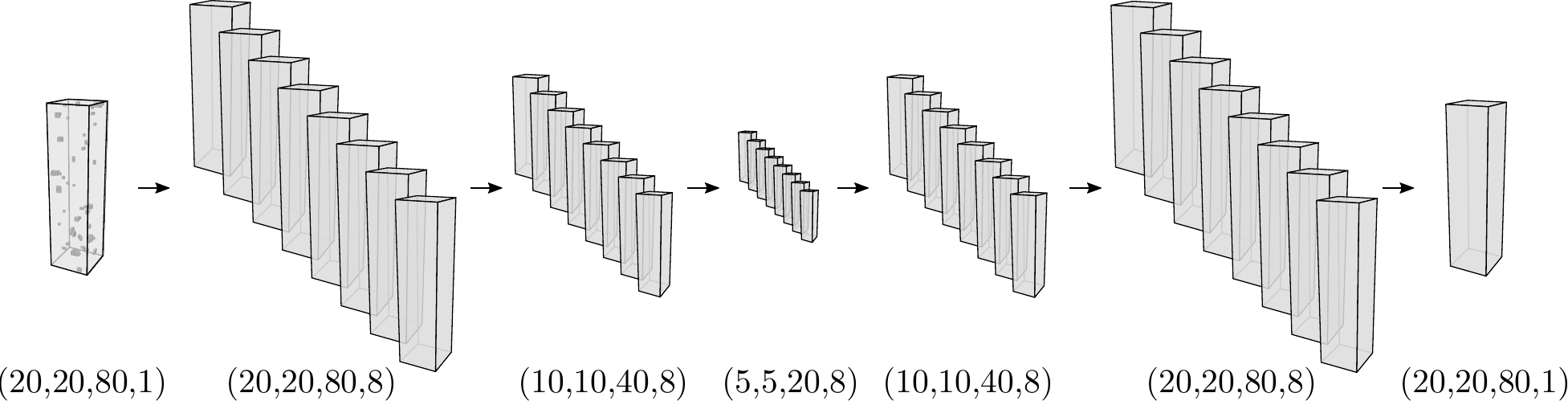}\\[2mm]
{\small
\begin{tabular}{l | l | l | l | l }
\hline
Deterministic         & Probabilistic         & Size         & Layer arguments  & Variables\\ \hline
Input                 & Input                 & -            & -  & -\\
Conv3D                & Convolution3DFlipout  & filters = 8  & kernel (3,3,3),  ReLU & 224/448\\
MaxPooling3D          & MaxPooling3D          & -            & kernel (2,2,2)  & 0\\
Conv3D                & Convolution3DFlipout  & filters = 8  & kernel (3,3,3),  ReLU & 1736/3472 \\
MaxPooling3D          & MaxPooling3D          & -            & kernel (2,2,2)  & 0\\
Conv3D                & Convolution3DFlipout  & filters = 8  & kernel (3,3,3),  ReLU & 1736/3472 \\
UpSampling3D          & UpSampling3D          & -            & size (2,2,2) & 0\\
Conv3D                & Convolution3DFlipout  & filters = 8  & kernel (3,3,3),  ReLU & 1736/3472\\
UpSampling3D          & UpSampling3D          & -            & size (2,2,2)  & 0\\
Conv3D                & Convolution3DFlipout  & filters = 8  & kernel (3,3,3),  ReLU & 1736/3472\\
Conv3D                & Convolution3DFlipout  & filters = 1  & kernel (3,3,3),  Linear & 217/434 \\
\hline
\end{tabular}
}
\caption{Illustration of the  encoder-decoder NN structure.
The convolutional layers are the only layers with trainable parameters and they could be either deterministic or probabilistic.
Padding: `same' is used for all the Conv3D, MaxPooling3D, and Convolution3DFlipout layers.}
\label{fig:NN-encoder-decoder}
\end{figure}

Therefore, starting with a ``linear'' KLE model and associated linear encoder-decoder with a latent variable space of dimensionality $\approx$12,000, we propose a non-linear heteroencoder with a total of $\approx$7,000 latent variables (refer to \fref{fig:NN-encoder-decoder}).
This heteroencoder CNN is parametrically parsimonious compared to fully connected feed-forward networks and CNNs that have all layers operate at full resolution, hence the $\approx$110,000 samples we have for training and validation are sufficient.

Training was carried out with a 7:1:2 train/validation/test split of the available data.
An Adam optimizer \cite{kingma2014adam} was employed with a learning rate of $10^{-3}$ over 300 epochs.
A batch size of 128 was used for the deterministic networks, while a larger batch size of 256 was used for the probabilistic networks (\sref{sec:BCNN}) to reduce the variance of the gradient estimations.
The weights were randomly initialized for deterministic networks, while a warm-start strategy is adopted for mean parameters of the Bayesian networks by initializing them from a corresponding trained, deterministic model.

\section{Evaluation of the convolutional neural network prediction performance} \label{sec:CNN_performance}

In the previous sections, transformations of the damage field and modifications of the MSE loss were motivated as strategies for improving the ability of the CNN to predict the relatively sparse regions displaying high damage.
We investigate these choices in a systematic manner to compare their effect on CNN performance and to reveal a more global picture of how the CNN model depends on these hyperparameters.

\subsection{Post-processing metrics}
While the MSE provides a standard measure of predictive accuracy for regression problems that is easily understood and formulated as a differentiable loss function, it does not completely capture some of the engineering aspects of performance.
Given the consequences of failure, it is more important to capture the locations of clusters of large damage values than it is to match the exact damage value predictions across the entire field.
Also predictions of failure in close proximity to the true failure locations are valuable in that they indicate clusters of porosity that are most susceptible to failure.
This motivates the consideration of alternative post-prediction performance metrics to evaluate the quality of the model after training.

\begin{figure}[h!]
\centering
\includegraphics[width=0.7\linewidth]{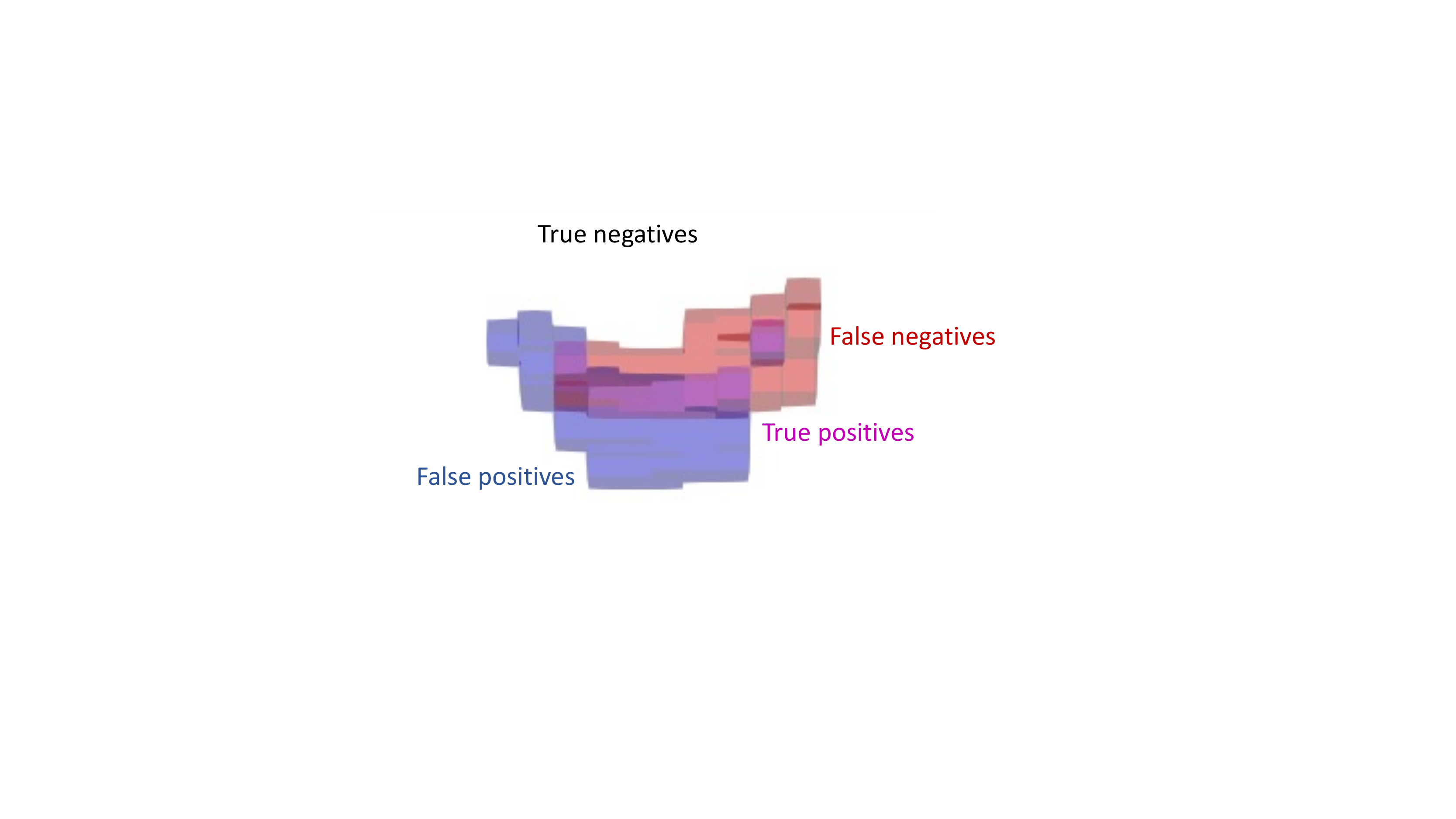}
\caption{Illustration of true positive voxels (magenta) where true (red) and predicted (blue) failure locations overlap. False positives are blue, false negatives are red, and true negatives are white.
}
\label{fig:overlap-metrics-diagram}
\end{figure}

\paragraph{Overlap metrics}
There are a number of metrics which provide a standard measure of similarity between two sets which appear frequently in binary classification tasks such as image segmentation and information retrieval.
Three such metrics, relevant to our work, are {\it precision}, {\it recall}, and {\it overlap} which are defined as:
\begin{align}
\text{Precision} &= \frac{TP}{TP+FP} \\
\text{Recall}    &= \frac{TP}{TP+FN} \\
\text{Overlap}   &= \frac{TP}{TP+FP+FN}
\end{align}
where $\text{TP}$, $\text{FP}$, and $\text{FN}$ denote true positives, false positives, and false negatives, respectively.
True positives are defined as the voxels which are predicted to fail by the CNN and also were actual failed voxels in the physical simulations that provided the data.
Set-wise: $\text{TP} = \text{Fail}_\text{true} \cap \text{Fail}_\text{pred}$.
These measure different aspects of similarity between true and predicted sets and are illustrated in \fref{fig:overlap-metrics-diagram}.
Note that in the present context these sets (and the associated spatial clusters) depend on a selected damage threshold value.

Intuitively, precision reflects the percentage of positive predictions which are accurate, while recall measures the percentage of true values which are captured by the prediction.
The F-score combines the precision and recall metrics to create an overall measure of accuracy from the harmonic mean of these two quantities.
On the other hand, overlap measures the extent to which the true and predicted regions coincide.
The overlap metric is also known as the Jaccard index, originally defined to measure the similarity of sets or, equivalently, binary vectors.
The Sorenson-Dice coefficient is a similar metric related to the Jaccard index by a monotonic transformation such that the two metrics induce the same ordering on sets and can be considered equivalent \cite{yeghiazaryan2018family}.
For our goal of predicting all the likely failure locations recall is the most important metric, since moderate over-prediction of failure provides a useful and conservative estimate of reliability.

\paragraph{Cluster overlap metric}
The overlap metrics based on voxel-wise comparison serve as strict measurements to evaluate the performance of the proposed CNN.
Considering the continuous nature of the damage field, the damage locations from CNN predictions often appear as spatially contiguous clusters for sufficiently high threshold values.
To have an evaluation of CNN performance that accommodates a sense of cluster proximity, we also calculate a relaxed overlap metric based on damage clusters instead of true and predicted damage values in the same voxel.
In this metric, we consider a predicted damage cluster as a true positive if it overlaps with a true damage cluster by at least one voxel.
Conversely, it is considered to be a false positive if it does not overlap with any true damage clusters.
Likewise, a true damage cluster is considered to be a false negative if it does not overlap with any predicted damage cluster.

\subsection{Cross-validation of model variants} \label{sec:cross-validation}
A 5-fold cross validation study of various choices of data transformation and loss weighting functions was carried out and evaluated according the the overlap metrics discussed in the previous section.
Variants of the CNN architecture shown in \fref{fig:NN-encoder-decoder} were trained for each combination of the following choices:
\begin{center}
\begin{tabular}{l | l | l | l}
\hline
\textbf{Data transformation} & Notation & \textbf{Loss re-weightings} & Notation \\
\hline
None & $\text{id}$ & None & $\text{id}$ \\
Softmax & $\sigma$ & Inverse histogram & $\text{IH}$ \\
Gaussian filter & $g$ & Inverse histogram per realization & $\text{IH}_{pr}$ \\
Softmax $\circ$ Gaussian filter & $t$ & &  \\
\hline
\end{tabular}
\end{center}

\begin{figure}[h!]
\centering
\subfloat[standard]{
\begin{minipage}{0.48\linewidth}
\includegraphics[width=0.95\linewidth]{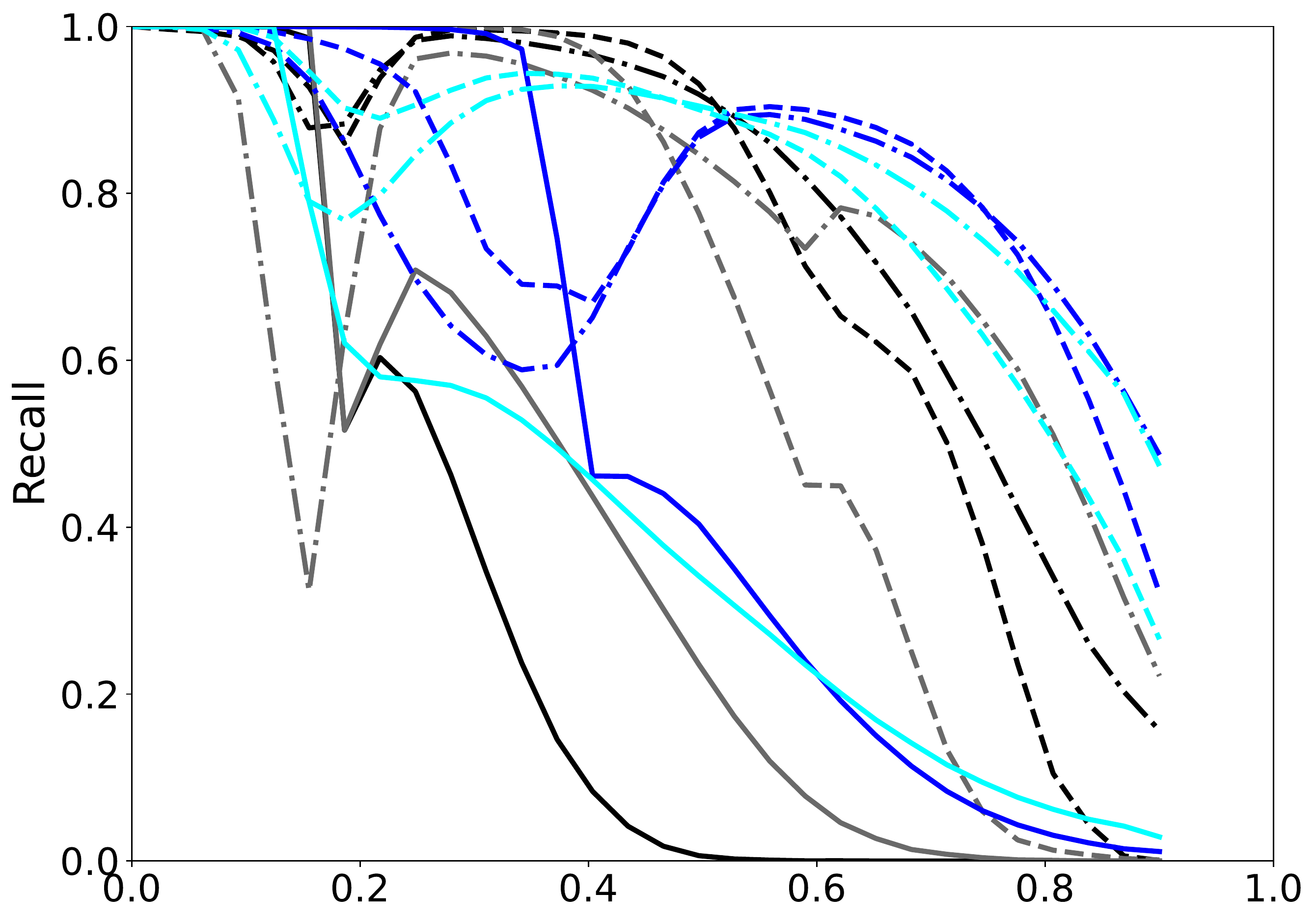}

\includegraphics[width=0.95\linewidth]{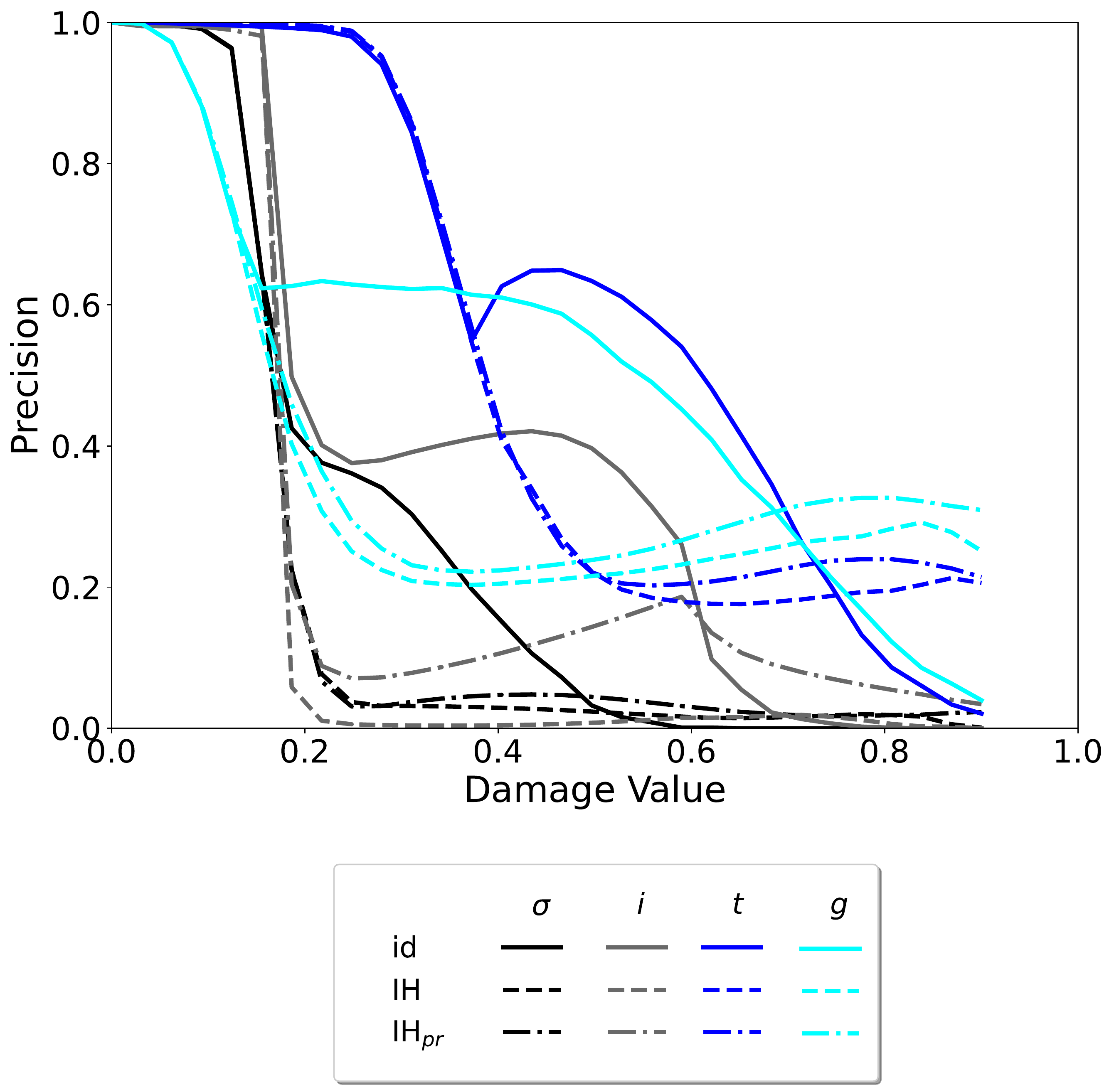}
\end{minipage}
}
\subfloat[cluster]{
\begin{minipage}{0.46\linewidth}
\includegraphics[width=0.95\linewidth]{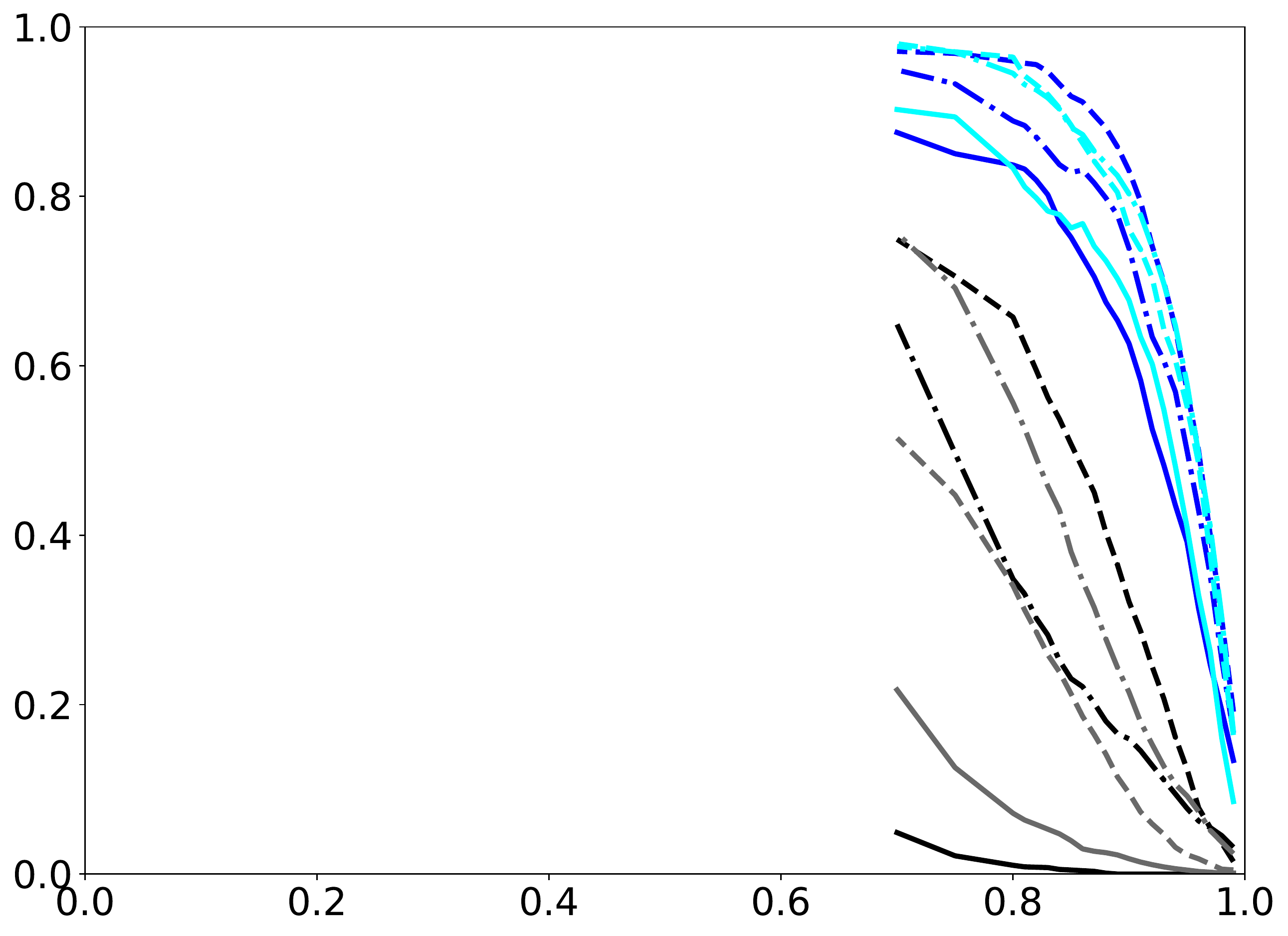}

\includegraphics[width=0.95\linewidth]{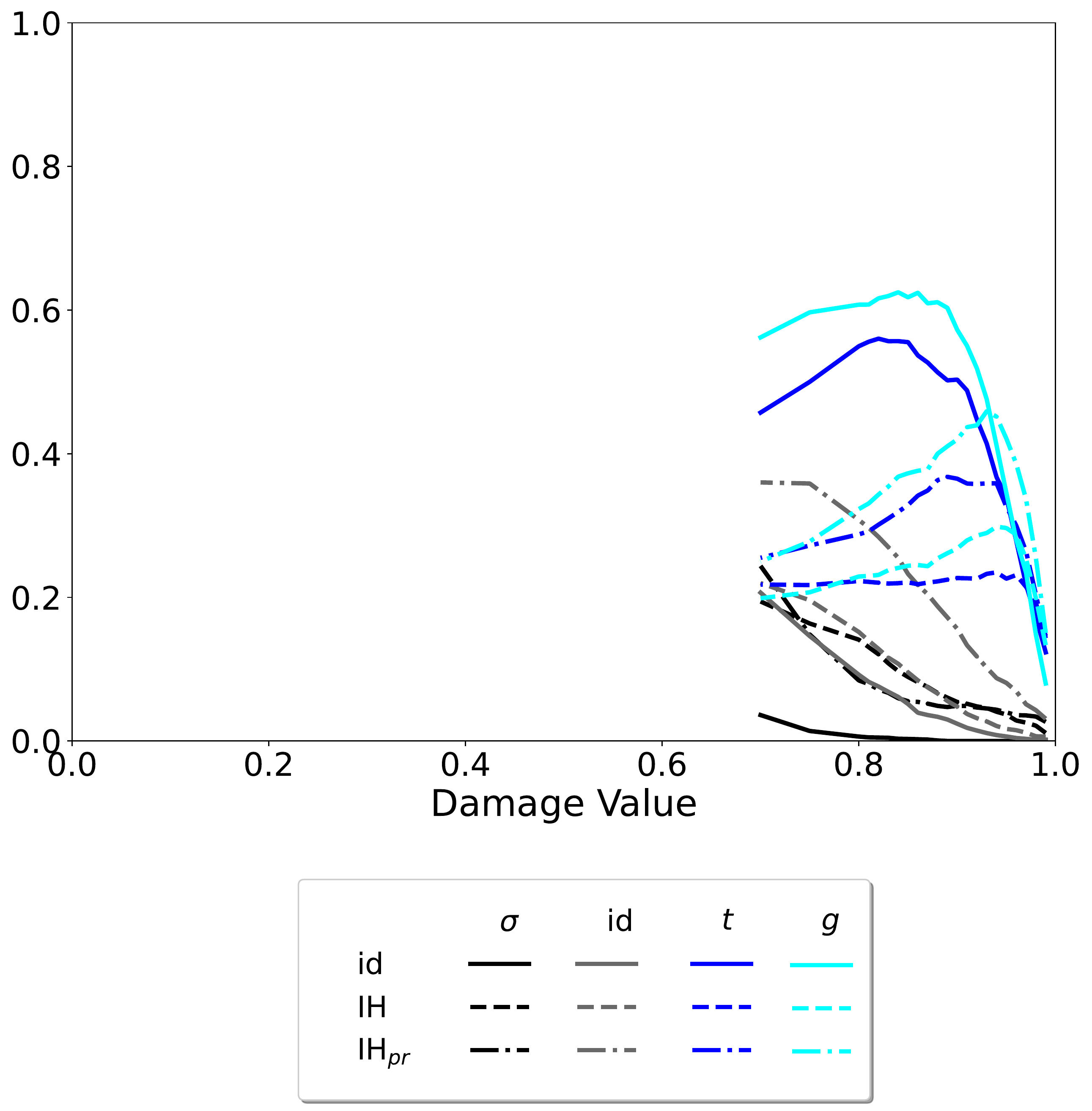}
\end{minipage}
}
\caption{Standard recall (left top) and precision (left bottom), and cluster overlap recall (right top) and precision (right bottom) for all 12 models. Color represents transformation while linestyle represents loss weighting.}
\label{fig:overlap}
\end{figure}

The precision and recall metrics for CNNs based on all 12 combinations are plotted as a function of damage value threshold according to the standard overlap metrics in  \fref{fig:overlap}a and according to the cluster overlap metric in \fref{fig:overlap}b.
Note we only show results for the cluster overlap metric for the range of interest, \ie where the threshold is high enough to discern distinct damage clusters and regions of likely failure, due to the computational expense of computing the cluster overlap metric.

Observe that recall and precision both improve as the data becomes increasingly smooth from the data transformation. In particular, the least smooth data is produced by softmax ($\sigma$) which tends to amplify gaps in the underlying field.
For the unweighted MSE, this displays the worst performance across the entire range of damage values.
The softmax and Gaussian filter transformation ($t$) and Gaussian filter ($g$) provide relatively smooth data the highest precision and recall values.
Both transformations are generally comparable and seem to marginally outperform one another in certain damage value regions.
This suggests that the smoothness induced by convolution is the most significant factor in improving CNN predictions on the damage field.

As expected, the loss weighting schemes combat class imbalance by emphasizing low-frequency, high-damage regions and result in an improvement to the recall and precision metrics.
Recall improves across the entire damage range while precision improves only at the upper end.
This is likely due to precision decreasing with false positives which increase with loss weightings as discussed in \sref{sec:weighted-loss}.

Note that the metrics display a general pattern of evolution.
Initially, there is a constant region at the maximal metric value of one.
This is followed by a region of rapid decrease which then slows and, possibly, reverses direction, sometimes forming a cusp in the graph.
After this, there is a local maximum and the function then monotonically decreases.

\begin{figure}[h!]
\centering
\includegraphics[width=1.0\linewidth]{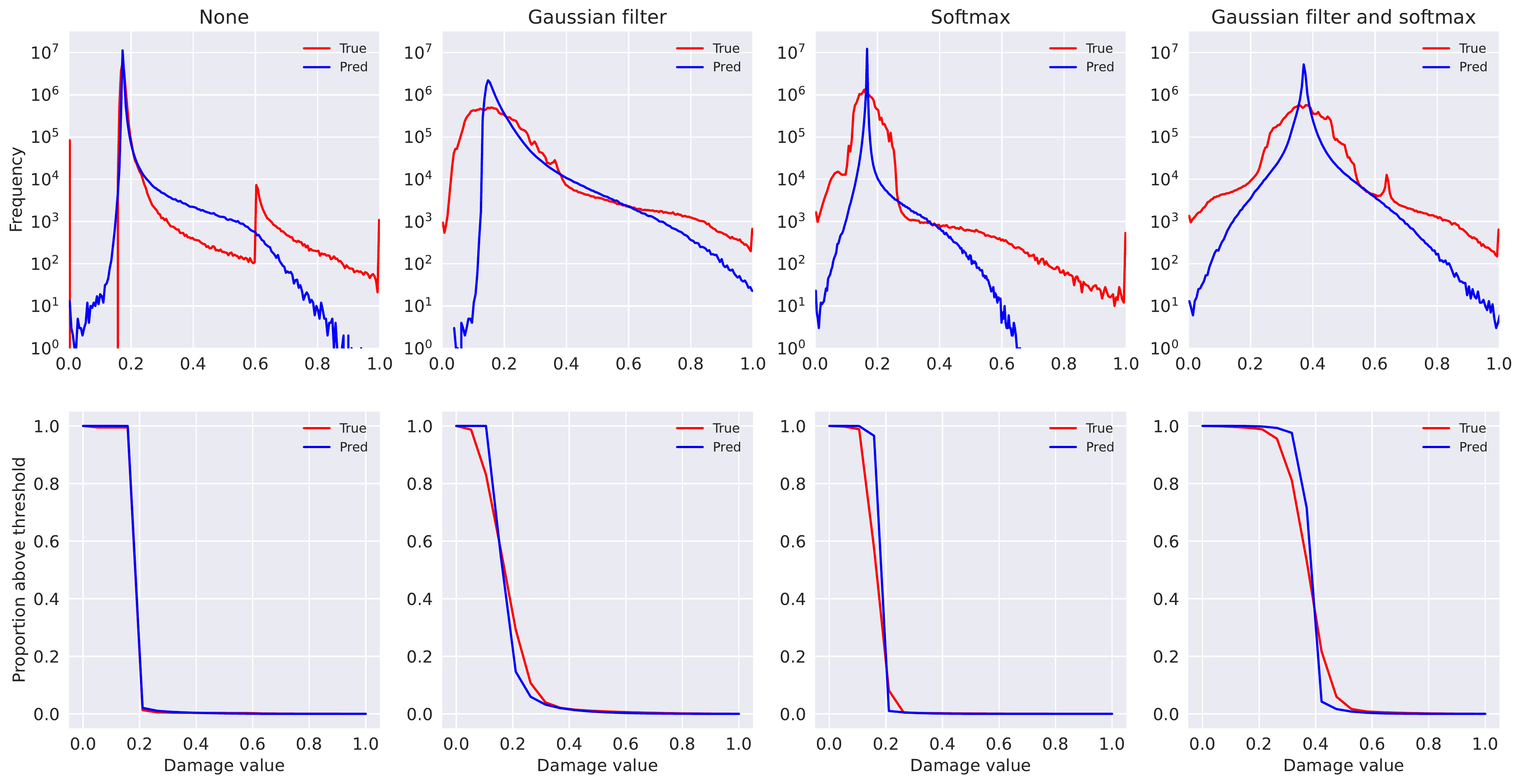}
\caption{Average damage frequency distributions for true and predicted fields (top row: no transformation, Gaussian filter, Softmax, both).
Superlevel set volume (fraction of voxels with damage above a threshold) as a function of damage threshold (bottom row).}
\label{fig:damage-distribution-analysis}
\end{figure}

To explain these phenomena, we consider \fref{fig:damage-distribution-analysis}.
For each data transformation, \fref{fig:damage-distribution-analysis} displays the damage value frequency distribution for true and predicted damage fields as well as the proportion of damage values above a given threshold for the unweighted MSE loss.
The initial plateaus are due to the prediction problem being degenerate for low thresholds, \ie essentially all voxels fail.
The cusps occur at the frequency peaks in the true damage fields suggesting that the trend reversal is because of the data frequency dependent performance of the CNN.
This is further supported by the observation that the loss re-weightings move the local maxima toward the higher end of the damage value region in \fref{fig:overlap}.

Note that the predicted frequency distribution does not exactly capture the shape of the true frequency distribution.
This discrepancy is amplified at the true damage field frequency peaks which are overemphasized by the predictions as an effect of class-imbalance.
This is reflected in the second row of \fref{fig:damage-distribution-analysis} which shows the volume of superlevel sets and reflects how quickly the damage fields transition to sparse clusters as the damage threshold increases.
Here, the superlevel sets reflect the collection of high damage clusters present after a thresholding operation transforms the continuous damage field to a binary field.
The predicted damage fields make a delayed, but faster transition to sparse clusters which  accounts for the initial rapid decrease in the overlap metrics since the two fields will temporarily display significantly different volumes.
This shows that the peaks of the true damage fields are smoother than those of the predicted fields resulting in different structure to the level set evolution.
Similar behavior is seen with each data transformation except the identity transformation where the transition of both the true and predicted level sets nearly coincide.

\subsection{Discussion of porosity sensitivity and predicted false positives} \label{sec:false_positives}

\begin{figure}[t!]
\centering
\includegraphics[width=0.22\linewidth]{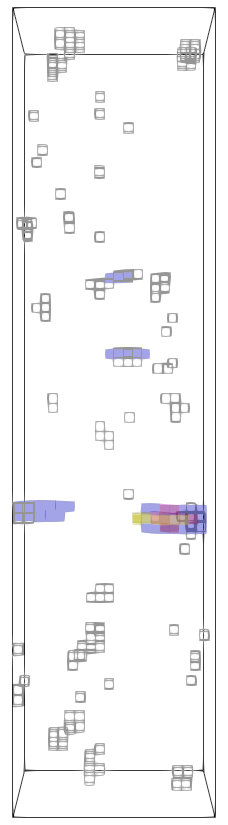}
\includegraphics[width=0.22\linewidth]{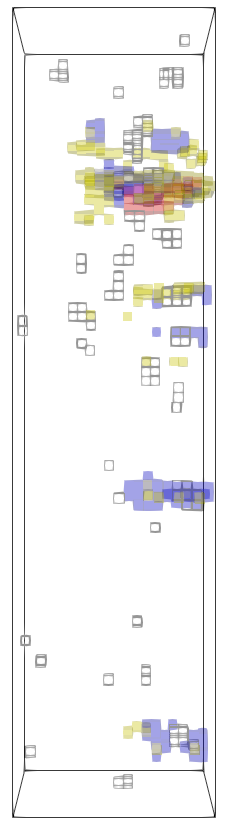}
\includegraphics[width=0.22\linewidth]{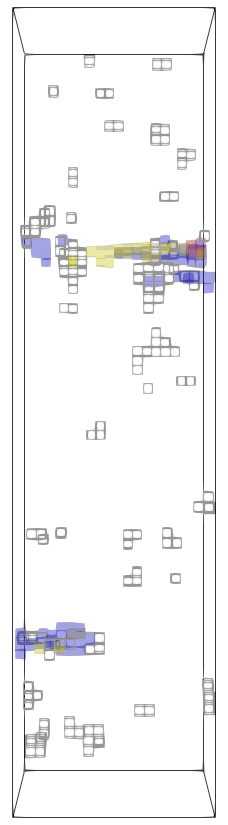}
\includegraphics[width=0.22\linewidth]{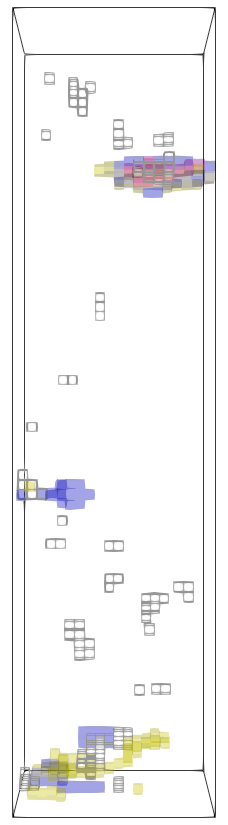}
\caption{CNN prediction of four selected realizations corresponding to  \fref{fig:sa-results}(a,g,h,l) for a threshold value of 0.8. Red: true damage, blue: CNN prediction, yellow: failure locations from the sensitivity analysis, gray wireframe: pore location. }
\label{fig:cnn-results-per-realization-weighting-sa}
\end{figure}

The CNN predicted results of realizations corresponding to those in \fref{fig:sa-results}(a,g,h,l), are presented in  \fref{fig:cnn-results-per-realization-weighting-sa}, where the pores, the true damage clusters and CNN predicted damage clusters for a threshold value of 0.8, and the failure locations from the sensitivity analysis are shown.
This CNN is trained with the softmax and Gaussian filter data transformation using an inverse histogram per realization weighted loss.
\fref{fig:cnn-results-per-realization-weighting-sa} shows that all the true damage clusters in these four realizations overlap with CNN predicted damage clusters.
This is consistent with the cluster recall curve in \fref{fig:overlap}, which has a value above 0.9 at a threshold of 0.8 for the selected data transformation and loss weighting.
Even though the cluster precision value in \fref{fig:overlap} for a threshold of 0.8 is around 0.3, we observe from \fref{fig:cnn-results-per-realization-weighting-sa} that some false positives indicate potential failure locations that were captured by the sensitivity analysis.
The CNN model appears to not only learn the complex nonlinear mapping between the pore distribution and damage initiation location (high recall score), but also learn subtle relationships between the two fields that were only revealed by the sensitivity analysis with expensive physics-based direct numerical simulations.

\section{A Bayesian convolutional model of localized failure} \label{sec:BCNN}
We investigated the addition of uncertainty quantification (UQ) to the hetereoencoder CNN model described in \sref{sec:CNN} using variational inference (VI) to approximate the Bayesian posterior distribution of the weights given training data.
This addition provides: (a) an assessment of the sufficiency of the training data volume as reflected in the distribution of uncertainty, and (b) a mitigation of possible overfitting by the regularization provided by Bayesian prior distributions.

\subsection{Uncertainty quantification with Bayesian inference}

In general, model predictions can be endowed with uncertainty estimates by allowing the model parameters to be random variables whose distributions are calibrated to the training data.
In the present context, we let $\bv{w}$ denote the parameters, both the weights and biases, of the neural network transformation, written as $\NN_{\bv{w}}(\porosityb)$, and let $\data = \{  \porosityb_i, \damageb_i \}$ denote the training data where $i\in\{1,N_s\}$ and $N_s$ is the number of data points.
The data is assumed to be generated by the neural network transformation and additive Gaussian noise such that $\damageb = \NN_{\bv{w}}(\porosityb)+\epsilonb=\preddamageb+\epsilonb$ where $\epsilonb \sim \mathcal{N}(\bv{0},\sigma^2)$ captures model discrepancy.
Note that we make a distinction between the exact and noisy outputs $\preddamageb$ and $\damageb$, respectively.
The noise accounts for the approximate nature of the model $\NN_{\bv{w}}$ and defines the likelihood distribution $p(\data \giv \bv{w})$ over the data given the weights.
The posterior probability of the parameters given the training data is provided by Bayes' rule:
\begin{equation} \label{eq:Bayes}
p(\bv{w} \giv \data) =  \frac{ p(\data \giv \bv{w}) p(\bv{w}) }{p(\data)} \ ,
\end{equation}
where $p(\bv{w})$ is the prior and $p(\data)$ is the evidence.
Uncertainty in the parameters as captured by $p(\bv{w} \giv \data)$ can then be pushed forward through the model to induce uncertainty in the outputs resulting in the pushed forward posterior distribution (PFP) \cite{sargsyan2015statistical}. We denote this as $p(\preddamageb \giv \porosityb, \data)$, which defines the distribution over outputs $\preddamageb$ given a particular input $\porosityb$. The PFP distribution is not analytically tractable so we employ a Monte Carlo scheme to estimate its first two moments based on sampling the weights $\bv{w} \sim p(\bv{w} \giv \data)$ according to the posterior.

\subsection{Approximating posteriors with variational inference} \label{sec:variational-inference}

The model evidence integral in the denominator of \eref{eq:Bayes} is typically intractable to evaluate, which necessitates methods to approximate the posterior \eqref{eq:Bayes}  and to sample from the PFP.
Briefly, there are a variety of techniques to enable Bayesian calibration/inference with a NN, \eg variational inference (VI)\cite{kingmaIntroductionVariationalAutoencoders2019}, stochastic gradient descent (SGD)-based \cite{mandt2017stochastic}, and Dropout \cite{gal2016dropout} schemes.
Variational inference minimizes a measure of discrepancy, such as Kullback-Liebler divergence
\begin{equation} \label{eq:vi-kl-cost}
q_{\bf{\thetab}} = \min_{q_{\bf{\thetab}} \in \mathcal{F}} \kldiv{q_{\bf{\thetab}}(\bv{w})}{p(\bv{w} \giv \data)} \ ,
\end{equation}
between a tractable parametric family $\mathcal{F}$ of probability distributions and the actual posterior $p(\bv{w} \giv \data)$ in order to select the optimal member $q_{\bf{\thetab}}$ of $\mathcal{F}$.
Here $\thetab$ is a vector parameterizing the family in the sense that $\mathcal{F} = \{ q_{\bf{\thetab}} \mid \thetab \in \Theta \subseteq \mathbb{R}^n \}$.
This gives us a closed form approximation of the true posterior whose fidelity is determined by the chosen family of distributions $\mathcal{F}$.
By sampling from this approximate posterior, we can form Monte Carlo estimates of the PFP distribution.

In this work, we use stochastic variational inference (SVI) which formulates a differentiable cost function equivalent to \eqref{eq:vi-kl-cost} and then carries out a form of gradient descent optimization.
The evidence lower-bound (ELBO) is a standard cost function in VI.
The ELBO $\mathcal{L}(\bv{\thetab})$ differs from the KL-divergence by a constant, the log of the evidence $\log p(\bv{\data})$:
\begin{equation}\label{eq:elbo-plus-kl}
\log p(\bv{\data}) = \mathcal{L}(\bv{\thetab}) + \kldiv{q_{\bv{\thetab}} (\bv{w})}{ p(\bv{w} \giv \data)} \ .
\end{equation}
A detailed derivation of \eref{eq:elbo-plus-kl} can be found in \cref{kingmaIntroductionVariationalAutoencoders2019}.
From \eref{eq:elbo-plus-kl}, it is clear that we can minimize the KL-divergence by maximizing the ELBO or, equivalently, minimizing the negative of the ELBO:
\begin{equation}\label{eq:elbo}
- \mathcal{L}(\bv{\thetab}) = \kldiv{q_{\bv{\thetab} } (\bv{w})}{p(\bv{w})} -  \ex_{q_{\bf{\thetab}} (\bv{w})}  \left[ \log p( \data \giv \bv{w}) \right] \ .
\end{equation}
Note that the ELBO no longer references the intractable posterior $p(\bv{w} \giv \data)$ and hence can be approximated in an optimization procedure.

To provide an understanding of the ELBO loss we relate it to the MSE loss used in deterministic CNN training, refer to \sref{sec:CNN}.
We first focus on the (second) data-dependent term of \eref{eq:elbo} and assume the likelihood $p( \data \giv \bv{w})$ takes the form of a mean field Gaussian with means given by the model predictions $\NN_{\bv{w}} (\porosityb_{s})$ and homoscedastic noise $\sigma$.
Note that we treat $\sigma$ as an additional optimization parameter that is calibrated along with $\thetab$ in the training of our Bayesian CNN models.
We can express the likelihood component as an expectation of the deterministic MSE
\begin{equation}\label{eq:elbo-likelihood-term}
-\ex_{q_{\bf{\thetab}} (\bv{w})}  \left[ \log p(\data \giv \bv{w}) \right]  = c + \frac{N_s N_v}{2} \log \sigma^2 + \frac{1}{2\sigma^2} \ex_{q_{\bf{\thetab}} (\bv{w})} \left[ \sum_{s=1}^{N_s} \norm{(\damageb_{s} - \NN_{\bv{w}} (\porosityb_{s}))}^2 \right]
\end{equation}
with respect to the approximate posterior $q_{\bf{\theta}} (\bv{w})$.
Note $N_v$ is the dimension of the damage field and $c$ is a constant with respect to parameters $\thetab$ and noise $\sigma^2$.
The (first) KL-divergence term in \eref{eq:elbo} involves the prior and exhibits no data-dependence.
It provides a regularization that drives the posterior toward the prior.
This suggests a connection between the ELBO cost and regularized least squares, similar to maximum likelihood estimation.

We can illustrate the connection by considering the case where the model is defined by $\NN_{\bv{W}} (\bv{x}) = \bv{W}\bv{x}$, \ie, the model is simply a linear transformation by the weight matrix $\bv{W} \in \mathbb{R}^{n \times n}$, as in PCA or a simple network model.
In this case, we can derive a closed form for the ELBO loss function.
Let $q_{\bf{\thetab}} (\bv{W})= \mathcal{N}(\bv{W} \giv \bm{\mu}_q, \bv{\Sigma}_q)$ and  $p(\bv{W}) = \mathcal{N}(\bv{W} \giv \bm{\mu}_p, \bv{\Sigma}_p)$ be matrix valued distributions denoting the approximate posterior and prior, respectively, where  $\bm{\mu}_q \in \mathbb{R}^{n \times n}$ is the matrix of means and $\bv{\Sigma}_q \in \mathbb{R}^{n^2 \times n^2}$ is the corresponding covariance.
Together, these constitute the variational parameters $\thetab = (\bm{\mu}_q,\bv{\Sigma}_q) \in \mathbb{R}^{n^2 + n^4}$.
Parameters $\bm{\mu}_p \in \mathbb{R}^{n \times n}$ and $\bv{\Sigma}_p \in \mathbb{R}^{n^2 \times n^2}$ are defined similarly for the prior $p(\bv{W})$ but are constants in the optimization process.
Given these assumptions, the ELBO has the form:
\begin{align}
-2\mathcal{L}_{\bv{\thetab}}
&=  \frac{1}{\sigma^2} \tr \{ (\bv{Y} -\bm{\mu}_q \bv{X})^T(\bv{Y} -\bm{\mu}_q \bv{X}) \} + (\bm{\mu}_p -\bm{\mu}_q)^T \bv{\Sigma}_p^{-1} (\bm{\mu}_p -\bm{\mu}_q) \label{eq:linear-elbo}\\
&+ \log \det(\bv{\Sigma}_q^{-1} \bv{\Sigma}_p) +\tr(\bv{\Sigma}_p^{-1} \bv{\Sigma}_q) + \frac{1}{\sigma^2}\tr \{ \bv{V} \bv{X} \bv{X}^T \} + N_s N_v \log \sigma^2 + c \ , \nonumber
\end{align}
where $\bv{X} = \left[ \porosityb_1 \cdots \porosityb_{N_s} \right]$ and $\bv{Y}=\left[ \damageb_1 \cdots \damageb_{N_s} \right]$ are data matrices consisting of the $N_s$ samples of porosity inputs and damage outputs, respectively.
The matrix $\Vb$ is defined by $V_{ij} = \Sigma_{kikj}$ given the definition $\mathbb{E}[W_{ij} W_{kl} ] = \Sigma_{ijkl} + \mu_{ij} \mu_{kl}$ with $\Sigma_{ijkl}$ being the components of $\Sigmab_q$.
Again, $c$ is a constant with respect to the variational parameters $\thetab$.
Note the details of this derivation, as well as a 1-dimensional example, are given in \aref{app:linear-elbo-derivation}.

Observe that with respect to the means, the problem has the form of least squares $\frac{1}{2\sigma^2} \tr \{ (\bv{Y} -\bm{\mu}_q \bv{X})^T(\bv{Y} -\bm{\mu}_q \bv{X}) \}$ with a quadratic regularization term $(\bm{\mu}_p -\bm{\mu}_q)^T \bv{\Sigma}_p^{-1} (\bm{\mu}_p -\bm{\mu}_q)$ arising from the KL-divergence $\kldiv{q_{\bv{\thetab}} (\bv{w} \giv \data)}{p(\bv{w})}$ portion of the ELBO.
The noise-dependent term $\log \det(\bv{\Sigma}_p \bv{\Sigma}_q^{-1}) +\tr(\bv{\Sigma}_p^{-1} \bv{\Sigma}_q)$ also arises from the KL-divergence and is a convex function attaining a unique minimum at $\bv{\Sigma}_q = \bv{\Sigma}_p$.
This term, along with the quadratic regularization, tend to drive the posterior to the prior.
The loss component $\frac{1}{2\sigma^2}\tr \{ \bv{V} \bv{X} \bv{X}^T \}$ arises from the likelihood portion of the ELBO and is characterized by the input data dependent matrix $\bv{X} \bv{X}^T$.
This term is uniquely minimized by $\bv{\Sigma}_q = \bv{0}$ (or when $\bv{V}$ is orthogonal to $\bv{X} \bv{X}^T$, when $\bv{X} \bv{X}^T$ is not full rank).
It has the effect of limiting the magnitude of the variance, which  minimizes the propagation of noise through the model $\bv{W}$ and improves the accuracy of predictions.

In the general, nonlinear case when the ELBO lacks a closed form expression, we can gain some intuition from \eqref{eq:elbo-likelihood-term} when the surrogate posterior is a Gaussian distribution $q_{\thetab}=\mathcal{N}(\bv{w} \giv \bm{\mu}_q,\bv{\Sigma}_q)$ by letting $\text{MSE}(\bv{w})=\frac{1}{N_s}\sum_{s=1}^{N_s} \norm{(\damageb_{s} - \NN_{\bv{w}} (\porosityb_{s}))}^2$ so that \eqref{eq:elbo-likelihood-term} becomes
\begin{equation}\label{eq:elbo-likelihood-conv}
-\ex_{q_{\bf{\thetab}} (\bv{w})}  \left[ \log p(\data \giv \bv{w}) \right]  = c + \frac{N_s N_v }{2} \log \sigma^2 + \frac{N_s}{2\sigma^2} (\mathcal{N}(\bv{w} \giv \bv{0},\bv{\Sigma}_q) \ast \text{MSE}(\bv{w}))(\bm{\mu}_q) \ ,
\end{equation}
where $(\mathcal{N}(\bv{w} \giv \bv{0},\bv{\Sigma}_q) \ast \text{MSE})(\bm{\mu}_q)=\int \mathcal{N}(\bv{w} - \bm{\mu}_q \giv \bv{0},\bv{\Sigma}_q) \text{MSE}(\bv{w}) \d{\bv{w}}$.
Hence the likelihood portion of the loss is a function of $\thetab = (\bm{\mu}_q,\bv{\Sigma}_q)$ and takes the form of a convolution in $\bm{\mu}_q$ with a Gaussian kernel of covariance $\bv{\Sigma}_q$.
This has the effect of smoothing the MSE to differing extents over different regions of parameter space. We expect smoothing to approximately preserve local minima so that they are carried over from the MSE to the ELBO.

\subsection{Training of the Bayesian neural network}
To train the model, we use SVI which utilizes gradient descent optimization of the ELBO cost function, \eref{eq:elbo}. At each iteration, a Monte Carlo scheme is used to estimate the gradient of the ELBO.
One way to obtain a gradient estimator is to reparametrize random variables to allow the gradient of the ELBO to be expressed as an expectation with respect to a probability density which can be estimated with standard Monte Carlo \cite{kingmaIntroductionVariationalAutoencoders2019}.
Due to the scale of neural networks, it is only feasible to use a small number of Monte Carlo samples.
This estimator is then subject to large variance, so we use {\it Flipout} \cite{wen2018flipout} to reduce the variance without having to use additional random samples.
The Flipout method decorrelates gradient samples within a batch by generating pseudo-independent perturbations of the samples via random matrices.

\subsection{Challenges with variational inference}
A common challenge in optimization is non-convexity of the loss function given a model that is non-linear in its parameters.
This leads to multiple local minima such that local optimizers may get stuck at a non-optimal solution, depending on initial conditions.
It then becomes necessary to employ strategies to lead the optimizer to a better local minimum such as initialization procedures or transformations of the loss function.
The form of the ELBO \eqref{eq:elbo} indicates that, while the KL-divergence term is convex for Gaussian distributions, the likelihood term in \eqref{eq:elbo-likelihood-term} will inherit non-convexity from the MSE loss.
Indeed, \crefs{kingmaIntroductionVariationalAutoencoders2019,bowmanGeneratingSentencesContinuous2016,sonderby2016train} report issues with SVI converging to poor local minima and employ strategies such as annealing to mitigate this.

In preliminary studies of the BCNN, we observed SVI converging to a model which predicted a uniform field at approximately the highest-frequency damage value independent of the input.
The uniform field solution was found to also be present in the deterministic CNN, which confirmed that the likelihood term does appear to inherit local optima from the MSE as suggested by \eqref{eq:elbo-likelihood-conv}.
By re-weighting the BCNN loss or using an informed initialization scheme, the uniform field local minimum could be avoided suggesting that the existence of this solution is a result of class imbalance.
Namely, the uniform field attains a low MSE value because the damage fields are approximately uniform in that the high-frequency damage values far outnumber other damage values.
Viewing the likelihood term in the form \eqref{eq:elbo-likelihood-conv} suggests that local optima in the likelihood will have mean values approximately equal to the weights representing local optima in the MSE.
Hence, to avoid poor local minima, we initialize the weight means $\bm{\mu}_q$ of the BCNN with the weights of the trained CNN in a ``warm-start'' scheme.

\section{Evaluation of the Bayesian convolutional neural network model predictions} \label{sec:BCNN_performance}

The BCNN provides a mean prediction, which can be thought of as a point estimate of the random weights and defines a deterministic mean transformation of $\porosityb$ to $\damageb$.
The variance of the weights induces a random distribution of outputs $\damageb$ around the mean providing uncertainty quantification on top of deterministic predictions.
Both the mean and the variance of the pushed forward posterior distribution over outputs can be estimated through Monte Carlo sampling of the pushed forward posterior distribution.
Each sample is generated by first sampling the weights according to their approximate Bayesian posterior \eqref{eq:Bayes} obtained through variational inference.
A forward pass with this random sample of the weights is used to generate a sample of the posterior distribution.
The sample mean and variance are then used to estimate the actual actual mean and variance.

\subsection{Mean predictions}
\begin{figure}[h!]
\centering
\includegraphics[width=0.45\linewidth]{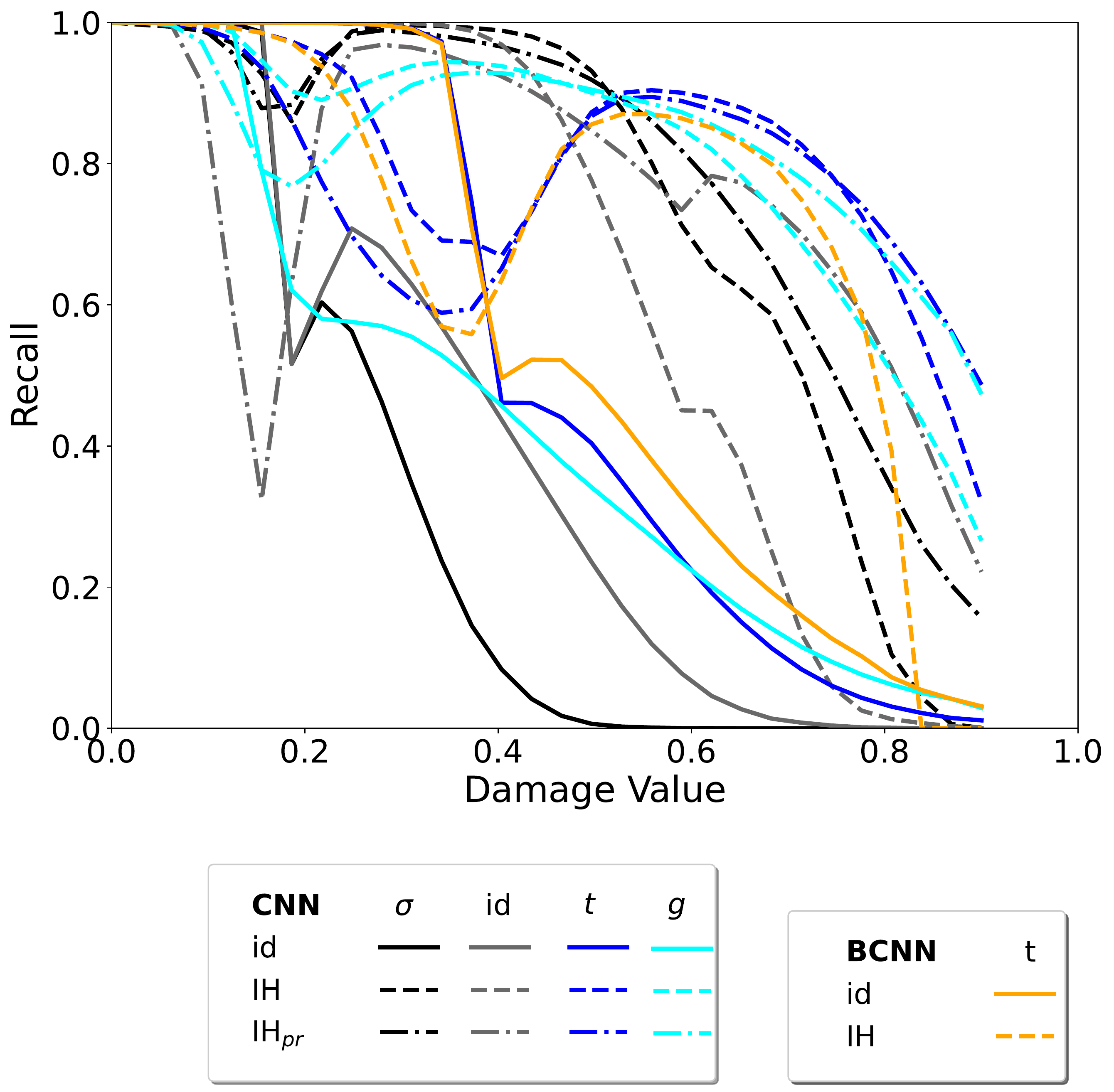}
\includegraphics[width=0.45\linewidth]{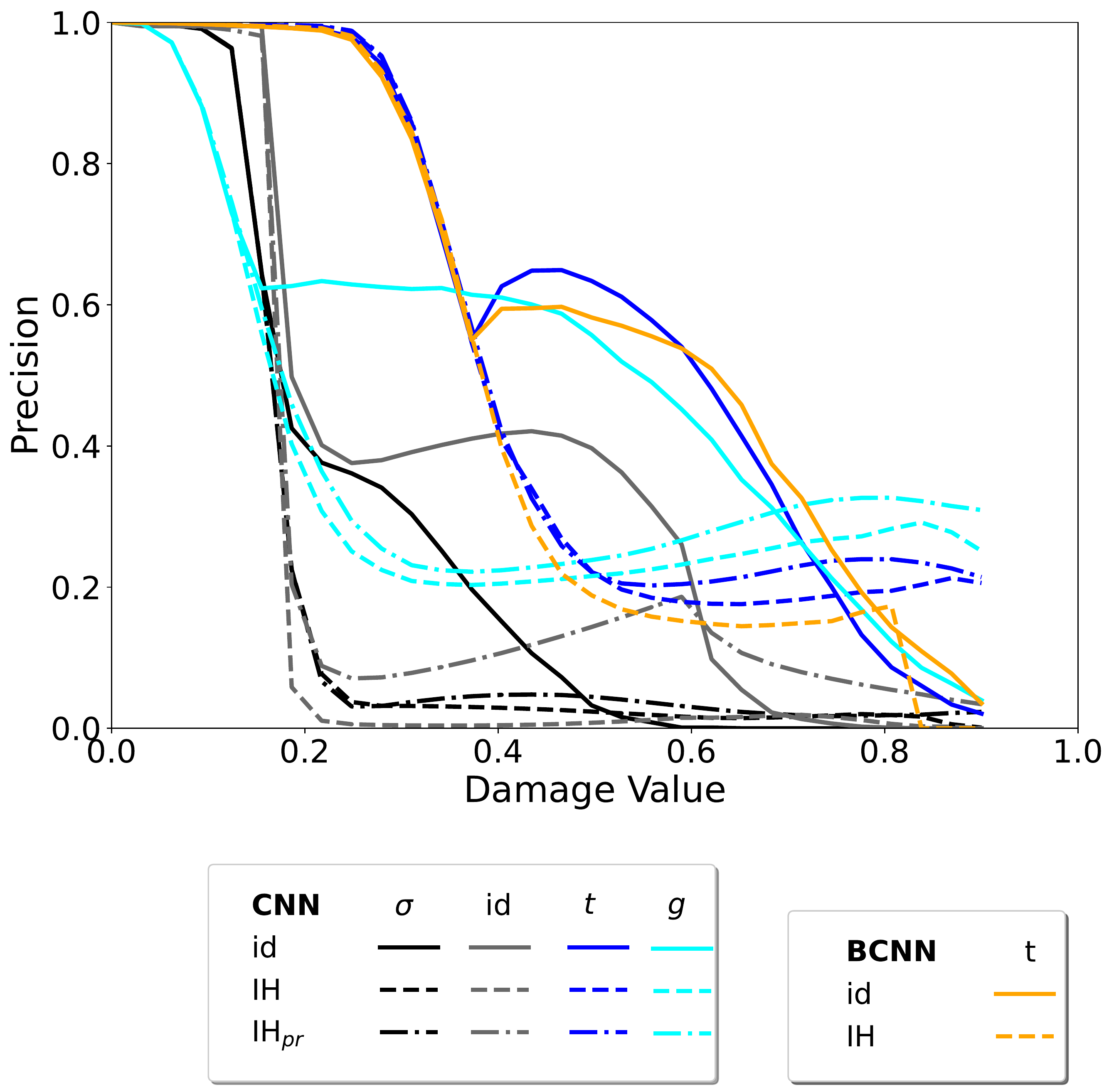}
\caption{Recall (left) and precision (right) of BCNN compared with CNN. Solid and dashed orange lines represent the unweighted and inverse histogram weighted BCNN, respectively.}
\label{fig:overlap-standard-bcnn}
\end{figure}

To analyze the mean predictions, the BCNN was compared to the cross-validation study from \sref{sec:CNN_performance} and evaluated according to the recall and precision metrics.
Two different likelihood loss weighting schemes were used analogous to the unweighted and the inverse histogram weighted versions of the MSE.
Given the results of \sref{sec:cross-validation}, only the softmax and Gaussian filter transformation were considered.
The results are displayed in \fref{fig:overlap-standard-bcnn} and show that for the unweighted loss, the mean predictions of the BCNN slightly outperform all of the deterministic CNNs.
This indicates that the inherent regularization provided by Bayesian neural networks is improving the quality of the solution found by the optimizer.
The BCNN trained using an inverse histogram weighting scheme for the likelihood term of the ELBO performs significantly worse than the best of the inverse histogram weighted CNNs.
For the unweighted loss, the calibrated noise standard deviation was $\sigma$ = 0.065, while for the weighted loss it was significantly higher.
This suggests that weighting schemes may have a different effect in the ELBO loss despite the connection to MSE.

\subsection{Uncertainty quantification} \label{sec:uq-posterior-predictive}

In the previous section, we established that the mean predictions of the unweighted BCNN perform similarly to the CNNs and even display improvement in the precision and recall metrics compared to the unweighted CNN.
In addition to the mean prediction, the BCNN also provides  useful information about uncertainty in the network predictions given the amount of training data.
In this section, the uncertainty in the posterior predictive distribution is analyzed and used to define a method to classify damage clusters according to their likelihood of failure.

\begin{figure}[h!]
\centering
\includegraphics[width=0.9\linewidth]{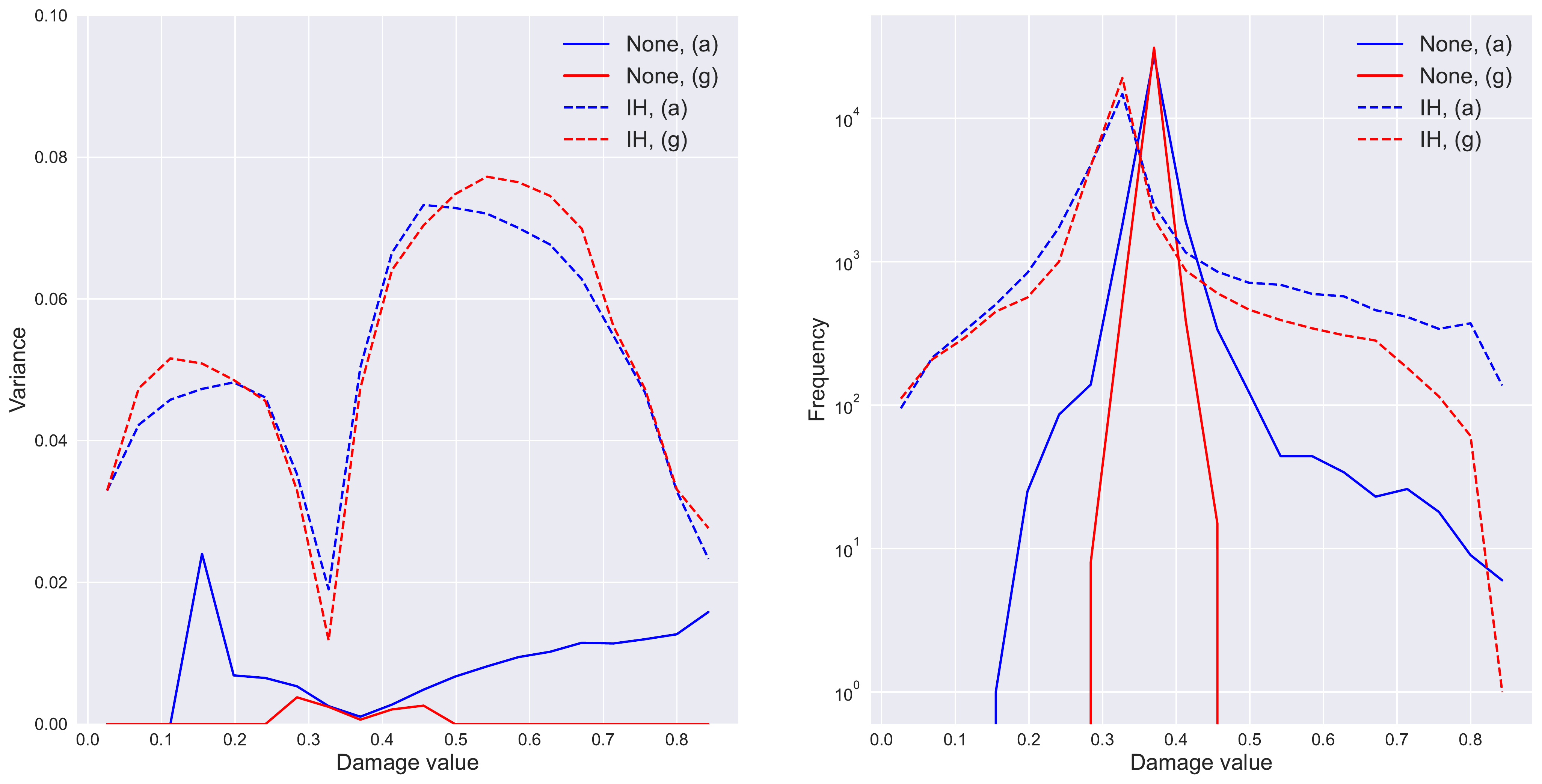}
\caption{Distribution of variance (left) and predicted damage value frequency distributions (right) for realizations (a) and (g).}
\label{fig:uncertainty-distribution}
\end{figure}

\paragraph{Uncertainty in the posterior predictive distribution}
The outputs of the BCNN are stochastic samples from a posterior predictive distribution which reflect the variance or uncertainty around the network's mean prediction.
A well-known phenomenon in Bayesian problems is the convergence of the posterior to a Dirac delta, a distribution with no uncertainty,  around a particular parameter value with increasing amounts of data.
Similarly, in simple regression problems with Bayesian networks, it can be seen that uncertainty in the outputs correlates with data sparsity in the domain, where sparse regions lead to higher uncertainty \cite{gelman1995bayesian}.
From these observations, we expect the uncertainty in the damage field BCNN to reflect data sparsity.
This suggests viewing the uncertainty as a function of damage value, similar to the loss distributions in the previous section.
This quantity is plotted in \fref{fig:uncertainty-distribution} where the variance across field voxels is binned according to the damage value.
Also displayed is the frequency distribution of the predicted damage field.

Examination of \fref{fig:uncertainty-distribution} indicates that the class imbalance aspect of the problem is affecting the uncertainty.
Comparison of the uncertainty distribution with the frequency distribution shows the damage value regions of high frequency exhibit a smaller level of uncertainty and vice versa.
Also note that the inverse histogram weighting of the log likelihood loss causes an overall increase in the uncertainty across the range of damage values.
In the limit of $w(\damageb)$ weighting any subset of data with zero weight, the size of our training set decreases.
Hence, down-weighting the high frequency regions of the damage spectrum is effectively reducing the amount of data we are using to train which results in a general increase in the amount of uncertainty.
This is consistent with what is expected in Bayesian inference.
Also observe that the uncertainty for the inverse histogram BCNN decreases toward larger damage values reflecting the emphasis that the inverse histogram weighting places on the high damage region.

\paragraph{Ranking damage clusters with UQ}
Now that we have a characterization of the uncertainty the BCNN predictions is capturing, we use both the mean and variance of the posterior predictive outputs to analyze the damage clusters.
A natural approach is to threshold the damage field to obtain multiple high damage value clusters (superlevel sets) and then classify them based on an appropriate metric representing the likelihood of failure.
To this end, we would like a metric which depends on the mean and variance of each cluster's Monte Carlo predictions.
A metric motivated by reliability analysis which satisfies this condition is the probability mass above a given threshold.
This allows ranking of the clusters according to their average damage value in a manner sensitive to the distribution around this average.
A cluster whose distribution shows a significant likelihood of producing a stochastic realization below its average value will be ranked below a cluster with the same average with lower variance.

\begin{figure}[h!]
\centering
\includegraphics[width=0.6\linewidth]{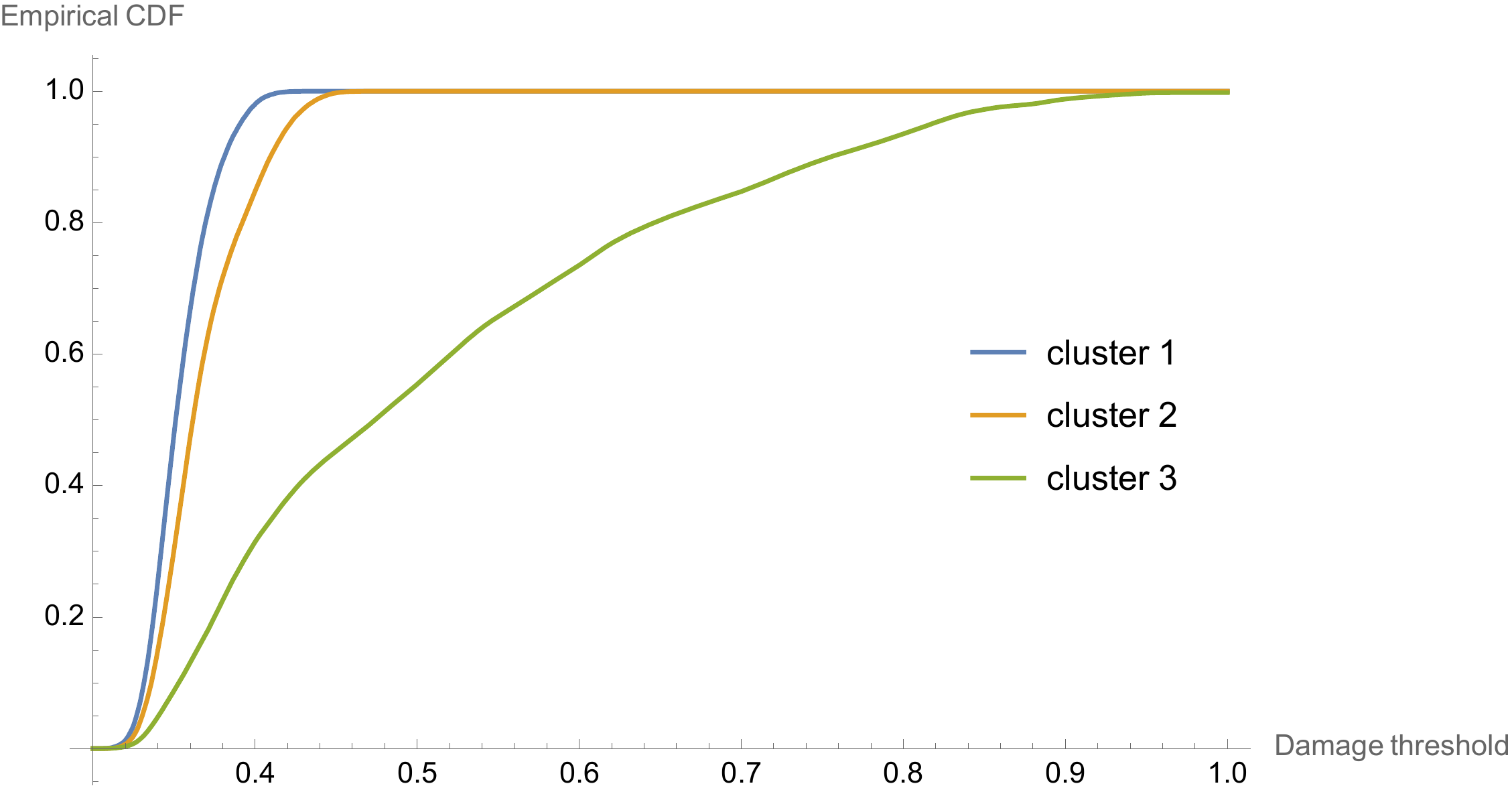}
\caption{Empirical CDFs for three different high damage clusters obtained after thresholding the damage field.}
\label{fig:damage-cluster-CDFs}
\end{figure}

\begin{figure}[h!]
\centering
\includegraphics[width=0.237\linewidth]{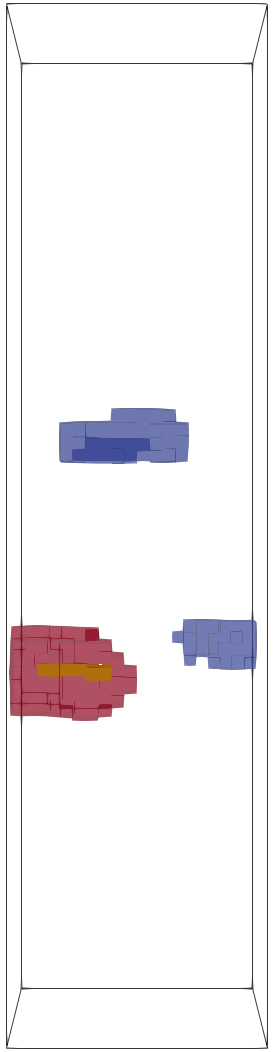}
\includegraphics[width=0.241\linewidth]{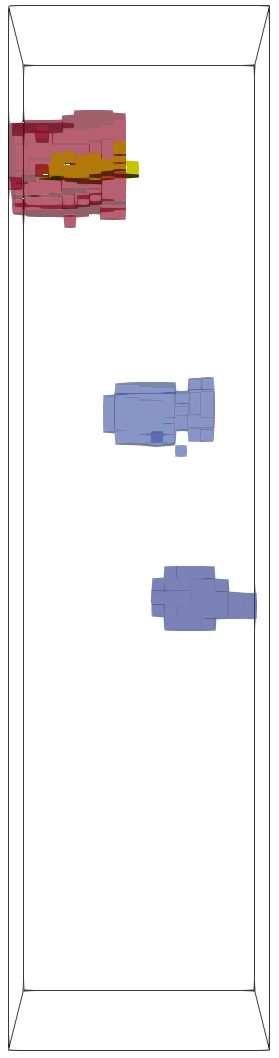}
\includegraphics[width=0.241\linewidth]{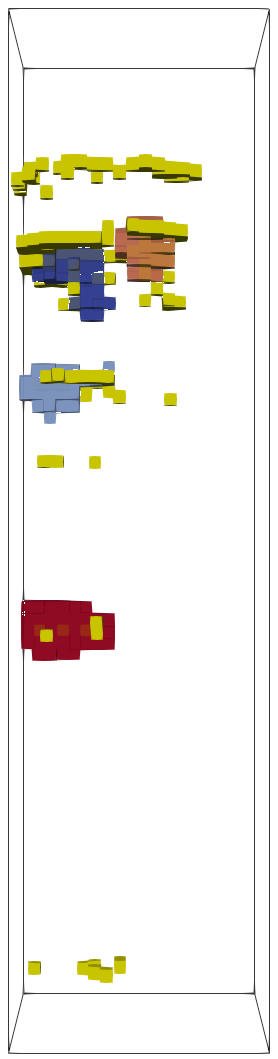}
\includegraphics[width=0.243\linewidth]{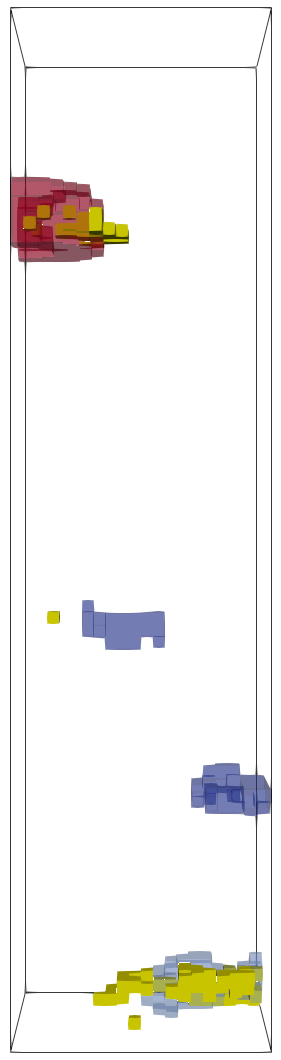}
\caption{Classification of damage clusters in realizations (a), (c), (g), and (j) using the probability mass above a threshold. Red indicates regions of higher mass while yellow represents the perturbed failure locations.}
\label{fig:cluster-rankings}
\end{figure}

To illustrate this concept the empirical CDFs of three clusters from the same damage field samples are shown in \fref{fig:damage-cluster-CDFs}.
The empirical distributions coincide at the lower and upper limits of the damage threshold but display a consistent ranking across an intermediate region of the damage threshold where the CDFs for all are less than one and greater than zero.
This suggests that the probability mass metric is fairly invariant with respect to choice of threshold and provides a robust method for ranking clusters.
This ranking scheme was carried out on realizations (a), (c), (g), and (j) from the sensitivity analysis in \fref{fig:sa-results} with results displayed in \fref{fig:cluster-rankings}.
Realizations (a), (c) display low sensitive and have only a single, small cluster of perturbed failures.
Observe that the probability mass ranking successfully determines the damage clusters containing these perturbed failures as the most likely failure regions.
Similarly, realization (j) has two perturbed failure regions and these clusters are ranked more highly than the clusters which do not coincide with perturbed failure locations.
Realization (g) is the most sensitive and each damage cluster coincides with a cluster of perturbed failures.
The probability mass ranking reflects this by assigning similar rankings to all four clusters indicating that the failure likelihoods are more difficult to distinguish in this case.

In this ranking scheme, a connected components algorithm \cite{pearce2005improved} designed for 3-dimensional fields was used.
It employs several strategies which provide optimal time complexity for the algorithm.
For more general clustering, we can represent voxel connectivity as a graph and use graph-based connected component algorithms such as depth-first-search or union-find.

\section{Discussion and conclusion} \label{sec:conclusion}

We demonstrated that a hetereoencoder comprised of an encoder of the input and a decoder that provides predictions of an output with different character is an effective representation of the porosity-to-damage-at-failure map for a porous ductile AM specimen.
The architecture proved useful in conjunction with data and loss transformations designed to alleviate the extreme class imbalance and spatial irregularity of the data, and yet preserve the accuracy of the predictions of failure locations.
We were also able to illustrate the utility of adding variational inference to this architecture.
Not only does it regularize the training/optimization of the network, VI provides indications of whether the data is sufficient to train the network beyond the usual comparison of training and test errors and more elaborate cross-validation.
The UQ provided by the BCNN architecture also proved useful in the pragmatic task of ranking the location most likely to fail.
The methods we developed have direct application to reliability assessment based on non-destructive CT scans and structural optimization given the inherent porosity of the AM process.

\begin{figure}[h!]
\centering
\includegraphics[width=1.0\linewidth]{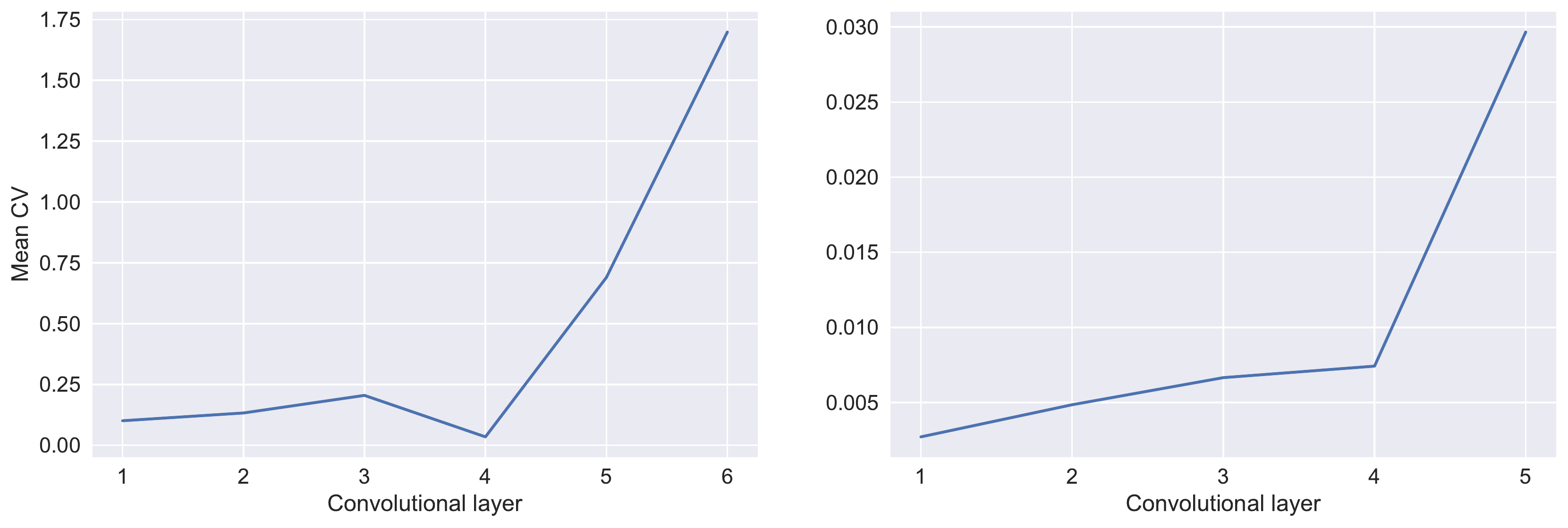}
\caption{Mean coefficient of variation (CV) of the kernel weights (left) and bias weights (right) as a function of layer.}
\label{fig:mean-cv-vs-layer}
\end{figure}

As a final observation to complement the examination of the uncertainty in the pushed forward posterior distribution of the BCNN in  \sref{sec:uq-posterior-predictive}, we studied the posterior distribution of the network parameters.
In particular, we looked at the overall uncertainty in the distribution of the weights as a function of each layer of the BCNN.
For each layer, we form a mean coefficient of variation (CV) from $\sum_i \sigma_i / \sum_i \mu_i$, the mean of standard deviations of the weights $\sigma_i$ over the mean of the means of the weights $\mu_i$.
\fref{fig:mean-cv-vs-layer} shows that on average, the uncertainty in both the convolutional kernel and bias parameters increases with depth in the network.
We offer two explanations for this phenomenon.
The first is that uncertainty in the initial layers of the network is propagated through a composition of nonlinear functions which may display high sensitivity to inputs.
Large uncertainty in the initial layers would lead to highly unpredictable outputs and thus, poor predictions.
Hence the uncertainty distribution may reflect effects of noise propagation.
The second explanation is based on the similarity of the BCNN architecture to an encoder-decoder which possesses a low-dimensional intermediate dimensional layer, which, in our case, is the result of the third convolutional layer of the stack of six.
Decoding transformations (layers 4-6) can be seen as allowing for larger variance as they are essentially adding information in reconstructing the output at the full resolution.

This observation has ramifications to network designs in that it may be sufficient to limit stochastic behavior to the final layer or layers of a neural network.
This would allow for UQ at significantly reduced computational cost.
It also motivates the study of how the uncertainty distribution across a network may correlate to the occurrence of different types of transformations such as the encoding and decoding portion of an autoencoder or heteroencoder.

The issue of choosing a threshold to predict regions likely to fail remains.
This suggests post-training tuning based on validation and other means of subverting or optimizing the threshold for accuracy.
A number of methods for segmenting images exist \cite{reddi1984optimal,wong1989gray,batenburg2008optimal,singla2017fast} and could be adapted to the failure prediction problem to improve the proposed methodology.

\section*{Acknowledgments}
TensorFlow and TensorFlow Probability ({\tt https://www.tensorflow.org/}) were used to construct and train the proposed neural network architectures.
The support of the Accelerated Strategic Computing (ASC) program at Sandia is gratefully acknowledged.
Sandia National Laboratories is a multimission laboratory managed and operated by National Technology and Engineering Solutions of Sandia, LLC., a wholly owned subsidiary of Honeywell International, Inc., for the U.S. Department of Energy's National Nuclear Security Administration under contract DE-NA-0003525.
The views expressed in the article do not necessarily represent the views of the U.S. Department of Energy or the United States Government.

\appendix
\section{Derivation of the ELBO for a linear model} \label{app:linear-elbo-derivation}
Herein, we provide supplementary details for derivation of the ELBO cost function \eref{eq:linear-elbo} associated with the mean field approximation of the parameter posterior for the linear transformation case described in \sref{sec:variational-inference}.
We subsequently provide the ELBO for the single parameter linear model case as a form of validation.

\subsection{Derivation for the {\it N}-dimensional case}
Recall that the cost function for variational inference takes the form of the negative ELBO as in \eref{eq:elbo}.
The KL-divergence term has a well-known analytical formula so we focus on deriving an expression for the likelihood term.
Starting with the mean-field normal form of the likelihood, we expand this term to
\begin{align*}
-\ex_{q_{\thetab}(\bv{W})} \left[ \log p(\data \giv \bv{W}) \right] &= \ex_{q_{\thetab}(\bv{W})} \left[ \log \prod_{i=1}^{N_s}  \mathcal{N}(\damageb_i \giv \NN_{\bv{W}}(\porosityb_i),\sigma^2 \bv{I})\right] \\
&= \frac{1}{2\sigma^2} \ex_{q_{\thetab}(\bv{W})} \left[ \sum_{i=1}^{N_s} \norm{\damageb_i - \NN_{\bv{W}}(\porosityb_i)}^2 \right] + \frac{N_sN_v}{2}\log \sigma^2 + c \\
&= \frac{1}{2\sigma^2} \ex_{q_{\thetab}(\bv{W})} \left[ \sum_{i=1}^{N_s} (\damageb_i - \bv{W}\porosityb_i)^T(\damageb_i - \bv{W}\porosityb_i) \right] + \frac{N_s N_v}{2}\log \sigma^2 + c \\
&= \frac{1}{2\sigma^2} \ex_{q_{\thetab}(\bv{W})} \left[ \tr\{ (\bv{Y}-\bv{W}\bv{X})^T(\bv{Y}-\bv{W}\bv{X}) \} \right] + \frac{N_sN_v}{2}\log \sigma^2 + c  \ ,
\end{align*}
where $\bv{Y} = \left[ \damageb_1 \cdots \damageb_{N_s} \right]$, $\bv{X} = \left[ \porosityb_1 \cdots \porosityb_{N_s} \right]$ are output and input data matrices.
In this case, the input-output transformation defined by the neural network model is linear such that $\NN_{\bv{W}}(\porosityb)=\bv{W}\porosityb$ with $\bv{W} \in \mathbb{R}^{n \times n}$ being parameters which we assume are normally distributed $\bv{W} \sim \mathcal{N}(\mub_q,\Sigmab_q)$. Note that we have assumed, without loss of generality, that the input and output vectors are of the same dimensionality $n=N_v$.
The last term $c = \frac{N_s N_v}{2}\log(2\pi)$ is a constant with respect to the variational parameters $\thetab$ and $\sigma^2$, which are both optimized during training.
Focusing on the matrix expression, we obtain
\begin{align*}
& \hspace{5mm} \frac{1}{2\sigma^2} \ex_{q_{\thetab}(\bv{W})} \left[ \tr\{ (\bv{Y}-\bv{W}\bv{X})^T(\bv{Y}-\bv{W}\bv{X}) \} \right] \\
&= \frac{1}{2\sigma^2} \ex_{q_{\thetab}(\bv{W})} \left[ \tr \{ \bv{Y}\bv{Y}^2 - \bv{Y}^T\bv{W}\bv{X} - \bv{X}^T\bv{W}^T\bv{Y} \} + \tr \{ \bv{X}^T\bv{W}^T\bv{W}\bv{X} \} \right] \\
&= \frac{1}{2\sigma^2}\tr\{ \bv{Y}^T\bv{Y}-\bv{Y}^T\mub_q\bv{X}-\bv{X}^T\mub_q^T\bv{Y} \} + \frac{1}{2\sigma^2}\tr\{ \ex_{q_{\thetab}(\bv{W})} \left[ \bv{W}^T\bv{W} \right] \bv{X}\bv{X}^T \} \ .
\end{align*}

To simplify notation, we will denote the covariance matrix by $\Sigmab_q \equiv \Sigmab$, where we have dropped the $q$ in anticipation of subscript indices. Typically, the covariance is thought of as a matrix $\Sigmab \in \mathbb{R}^{n^2 \times n^2}$ over the random vector $\text{vec}(\mub_q) \in \mathbb{R}^{n^2}$ where $\text{vec}(\cdot)$ is the operation which maps an $n \times m$ matrix into a column vector in $\mathbb{R}^{nm}$. Here, we adopt the alternative perspective of defining the covariance as a rank-4 tensor
\begin{equation*}
\Sigmab = \ex_{q_{\thetab}(\bv{W})} \left[ (\bv{W}-\mub_q) \otimes (\bv{W}-\mub_q) \right] = \ex_{q_{\thetab}(\bv{W})} \left[ \bv{W} \otimes \bv{W} \right] - \mub_q \otimes \mub_q
\end{equation*}
so that we have $\ex_{q_{\thetab}(\bv{W})}\left[ \bv{W} \otimes \bv{W} \right]=\Sigmab + \mub_q \otimes \mub_q$ or, equivalently, in component notation, $\ex_{q_{\thetab}(\bv{W})}\left[ W_{ij}W_{jk} \right] = \Sigma_{ijkl} + \mu_{ij}\mu_{kl}$.
To obtain $\ex_{q_{\thetab}(\bv{W})}\left[ \bv{W}^T\bv{W} \right]$, we can contract this expression, with implied Einstein summation convention, to get
\begin{equation*}
\left[ \ex_{q_{\thetab}(\bv{W})}\left[ \bv{W}^T\bv{W} \right] \right]_{ij} = \Sigma_{kikj} + \mu_{ki}\mu_{kj} = V_{ij} + \mu_{ki}\mu_{kj}
\end{equation*}
where we have defined $V_{ij} = \Sigma_{kikj}$.
Returning to matrix notation, we have
\begin{align*}
\frac{1}{2\sigma^2}\tr\{ \ex_{q_{\thetab}(\bv{W})} \left[ \bv{W}^T\bv{W} \right] \bv{X}\bv{X}^T \} &= \frac{1}{2\sigma^2} \tr\{ (\bv{V} + \mub_q^T \mub_q) \bv{X}\bv{X}^T \} \\
&= \frac{1}{2\sigma^2}\tr\{ \bv{X}^T\mub_q^T\mub_q\bv{X} \} + \frac{1}{2\sigma^2}\tr\{ \bv{V}\bv{X}\bv{X}^T \}
\end{align*}
so that the likelihood term can finally be written in terms of known matrix quantities
\begin{align*}
-\ex_{q_{\thetab}(\bv{W})} \left[ \log p(\data \giv \bv{W}) \right]  &= \frac{1}{2\sigma^2}\tr\{ \bv{Y}^T\bv{Y}-\bv{Y}^T\mub_q\bv{X}-\bv{X}^T\mub_q^T\bv{Y} + \bv{X}^T\mub_q^T\mub_q\bv{X} \} \\
&+ \frac{1}{2\sigma^2}\tr\{ \bv{V}\bv{X}\bv{X}^T \} + \frac{N_sN_v}{2}\log \sigma^2 + c \\
&= \frac{1}{2\sigma^2} \tr\{ (\bv{Y}-\mub_q\bv{X})^T(\bv{Y}-\mub_q\bv{X})\}+\frac{1}{2\sigma^2}\tr\{ \bv{V}\bv{X}\bv{X}^T \} \\
&+ \frac{N_sN_v}{2}\log \sigma^2 + c
\end{align*}
Hence, our full negative ELBO cost function is
\begin{align*}
-2\mathcal{L}_{\bv{\thetab}}
&=  \frac{1}{\sigma^2} \tr \{ (\bv{Y} -\bm{\mu}_q \bv{X})^T(\bv{Y} -\bm{\mu}_q \bv{X}) \} + (\bm{\mu}_p -\bm{\mu}_q)^T \bv{\Sigma}_p^{-1} (\bm{\mu}_p -\bm{\mu}_q) \\
&+ \log \det(\bv{\Sigma}_q^{-1} \bv{\Sigma}_p) +\tr(\bv{\Sigma}_p^{-1} \bv{\Sigma}_q) + \frac{1}{\sigma^2}\tr \{ \bv{V} \bv{X} \bv{X}^T \} + N_s N_v \log \sigma^2 + c \ .
\end{align*}
as in \eref{eq:linear-elbo}.

\subsection{Reduction to the 1-dimensional case}
To gain intuition into how the various terms contribute to the loss function, consider the case where $\bv{W}=w \in \mathbb{R}^{1 \times 1}$, \ie, our matrix is just scaling a univariate input by a constant to produce a scalar output.
In this single parameter case, the negative ELBO cost function becomes
\begin{align*}
-\mathcal{L}_{\bv{\thetab}} &= \frac{1}{2\sigma^2} \sum_{i=1}^{N_s} (\damage_i - \mu_q\porosity_i)^2 + \frac{1}{2\sigma_p^2}(\mu_p - \mu_q)^2 \\
&+ \frac{1}{2}\left( \log\frac{\sigma_p^2}{\sigma_q^2} + \frac{\sigma_q^2}{\sigma_p^2} \right) + \frac{\sigma_q^2}{2\sigma^2}\sum_{i=1}^{N_s}\porosity_i^2  + \frac{N_s}{2}\log \sigma^2 + c \ ,
\end{align*}
where $q_{\thetab}(w)=\mathcal{N}(w \giv \mu_q, \sigma_q^2)$, $p(w)=\mathcal{N}(w \giv \mu_p, \sigma_p^2)$ are the surrogate posterior and the prior on $w$, respectively.

To check the correctness of this formula, we show that we can express the negative ELBO in the 1-dimensional case as $L(\mu_l,\sigma_l,\mu_p,\sigma_p,\mu_q,\sigma_q)$, where $\mu_l$, $\sigma_l$ are the mean and standard deviation of the likelihood probability density function of parameter $w$.
Bayesian inference does not distinguish between the prior and likelihood in the sense that $p(w \giv \data) =  \frac{ p(\data \giv w) p(w) }{p(\data)}=\frac{  p(w) p(\data \giv w) }{p(\data)}$.
Hence, as VI will find the exact Gaussian posterior in this case, we should expect that the optimal solution $\mu_q^*,\sigma_q^*$ is invariant under exchanging the prior and likelihood parameters (mean and standard deviation), \ie,
\begin{equation*}
\mu_q^*,\sigma_q^* = \argmin_{\mu_q,\sigma_q} L(\mu_l,\sigma_l,\mu_p,\sigma_p,\mu_q,\sigma_q) = \argmin_{\mu_q,\sigma_q} L(\mu_p,\sigma_p,\mu_l,\sigma_l,\mu_q,\sigma_q)
\end{equation*}

Letting $\bv{y} = \left[ \damage_1 \cdots \damage_{N_s} \right]$, $\bv{x} = \left[ \porosity_1 \cdots \porosity_{N_s} \right]$, recall that Bayesian linear regression with a uniform prior is defined by the log likelihood $\chi^2 = -\frac{1}{2\sigma^2}(\bv{y}-w\bv{x})^T(\bv{y}-w\bv{x})$.
We can extract the likelihood mean and variance from the first and second order derivatives of the log likelihood \cite{sivia2006data}:
\begin{align*}
\frac{\partial}{\partial w} \chi^2 = 0 &\Leftrightarrow w = \frac{\bv{y}^T \bv{x}}{\bv{x}^T\bv{x}} \ , \\
\frac{\partial^2}{\partial w^2}\chi^2 &= -\frac{\bv{x}^T\bv{x}}{\sigma^2} \ ,
\end{align*}
so that $\mu_l = \frac{\bv{y}^T\bv{x}}{\bv{x}^T\bv{x}}$ and $\sigma_l^2=\frac{\sigma^2}{\bv{x}^T\bv{x}}$. Based on these expressions, the first term in the negative ELBO can be written as
\begin{align*}
\frac{1}{2\sigma^2} \sum_{i=1}^{N_s} (\damage_i - \mu_q\porosity_i)^2 &= \frac{1}{2\sigma^2}(\bv{y}-\mu_q \bv{x})^T(\bv{y}-\mu_q \bv{x}) \\
&= \frac{1}{2\sigma^2}\bv{x}^T\bv{x}\left( \frac{\bv{y}^T\bv{y}}{\bv{x}^T\bv{x}} - 2 \mu_q \frac{\bv{y}T\bv{x}}{\bv{x}^T\bv{x}} + \mu_q^2 \right) \\
&= \frac{1}{2\sigma_l^2}\left( \frac{\bv{y}^T\bv{y}}{\bv{x}^T\bv{x}} - 2 \mu_q \mu_l + \mu_q^2 \right) \\
&= \frac{1}{\sigma_l^2}(\mu_q-\mu_l)^2 + \frac{1}{2\sigma_l^2}\left( \frac{\bv{y}^T\bv{y}}{\bv{x}^T\bv{x}}-\mu_l^2 \right) \ ,
\end{align*}
in which the first term is a complete square of the difference between the likelihood and variational posterior means.
We also note that
\begin{equation*}
\frac{\sigma_q^2}{\sigma^2}\sum_{i=1}^{N_s}\porosity_i^2 =
\frac{\sigma_q^2}{\sigma^2}\bv{x}^T\bv{x} =  \frac{\sigma_q^2}{\sigma_l^2} \ .
\end{equation*}
Hence the univariate ELBO can be re-expressed as
\begin{align*}
L(\mu_l,\sigma_l,\mu_p,\sigma_p,\mu_q,\sigma_q) &= \frac{1}{2\sigma_l^2}(\mu_q-\mu_l)^2 + \frac{1}{2\sigma_p^2}(\mu_p-\mu_q)^2 \\
&+ \frac{1}{2}\left(\frac{\sigma_q^2}{\sigma_p^2} + \frac{\sigma_q^2}{\sigma_l^2} - \log \sigma_q^2 \right)+ c'  \ ,
\end{align*}
where $c' = \frac{N_s}{2}\log \sigma^2 + \frac{1}{2\sigma_l^2}\left( \frac{\bv{y}^T\bv{y}}{\bv{x}^T\bv{x}}-\mu_l^2 \right) + \frac{1}{2}\log \sigma_p^2$ is a constant with respect to the variational parameters $\mu_q,\sigma_q$.
Therefore, the optimal solution is
\begin{eqnarray*}
\sigma_q^{*2} &=& \left( \sigma_p^{-2}+\sigma_l^{-2} \right)^{-2} \\
\mu_q^* &=& \sigma_q^{*-2} \left(  \sigma_p^{-2} \mu_p+\sigma_l^{-2} \mu_l \right)
\end{eqnarray*}
which is preserved under the interchange of the prior and likelihood parameters, as expected.



\end{document}